\documentclass[12pt,oneside]{article}

\renewcommand{\title}[1]{\null\vspace{25mm}\noindent{\Large{\bf #1}}\vspace{10mm}}
\newcommand{\authors}[1]{\noindent{\large #1}\vspace{20mm}}
\newcommand{\address}[1]{{\center{\noindent #1\vspace{10mm}}}}
\renewcommand{\abstract}[1]{\vspace{17mm}
\noindent{\small{\em Abstract.} #1}\vspace{2mm}}     
 
\newcommand{\be}{\begin{equation}}
\newcommand{\ee}{\end{equation}}
\newcommand{\ba}{\begin{array}}
\newcommand{\ea}{\end{array}}
\newcommand{\bea}{\begin{eqnarray}}
\newcommand{\eea}{\end{eqnarray}}

\newcommand{\nm}{\nonumber}

\newcommand{\br}{{\bf R}^{4}}

\newcommand{\Li}{{\cal L}_\tau}
\newcommand{\del}{\delta_\tau}
\newcommand{\id}{i_\tau}
\newcommand{\g}{g(\tau)}
\newcommand{\tr}{{\rm tr \, }}

\newcommand{\intM}[1]{\int_{{\cal M}_#1}}
\newcommand{\intR}[1]{\int_{{\bf R}^{#1}}}

\newcommand{\SLT}{{\cal S}}
\newcommand{\ds}{\displaystyle}
\newcommand{\W}[2]{\ds{\frac{\delta #1}{\delta #2}}}
\newcommand{\ST}[2]{\W{\Sigma}{#1}\W{\Sigma}{#2}}
\newcommand{\LST}[2]{\W{\Sigma}{#1}\W{}{#2}+\W{\Sigma}{#2}\W{}{#1}}
\newcommand{\WI}{{\cal W}}

\newcommand{\bh}{\frac{1}{2}}

\newcounter{saveeqn}

\newcommand{\captionit}[1]{\caption{\small\sl{#1}}}

\begin{document}   \setcounter{table}{0}
 
\begin{titlepage}
\begin{center}
\hspace*{\fill}{{\normalsize \begin{tabular}{l}
                         {\sf LYCEN 2001-81}\\
                              {\sf REF. TUW 01-21}\\
                              {\sf November 2001}\\
                              %{\sf\footnotesize PAYS: 11.10.Kk, 11.15} \\
                              \end{tabular}   }}

%\medskip

\title{Symmetries  of Topological Field Theories\\
in the BV-framework}

\medskip 

\authors{
F.~Gieres$^{\, a)}$,  
%\footnote{email: gieres@ipnl.in2p3.fr}, 
J.M.~Grimstrup$^{\, b)}$\footnote{Work 
supported by ``The Danish Research Academy''.}, 
H.~Nieder$^{\, c)}$, 
T.~Pisar$^{\, b)}$\footnote{Work supported 
by the ``Fonds zur F\"orderung der
Wissenschaflichen Forschung" under Project Grant Number P11582-PHY.}, 
M.~Schweda$^{\, b)}$\footnote{email: mschweda@tph.tuwien.ac.at}
}

%\vspace{-20mm} 

%\renewcommand{\thefootnote}{\arabic{footnote}}       
%\addtocounter{footnote}{-5}

\address{$^{a)}$ Institut de Physique Nucl\'eaire, 
Universit\'e Claude Bernard \\
43, boulevard du 11 novembre 1918 \\
      F - 69622 - Villeurbanne (France)}

\vspace{-5mm}

\address{$^{b)}$ Institut f\"ur Theoretische Physik, 
Technische Universit\"at Wien\\
      Wiedner Hauptstra\ss e 8-10 \\
      A - 1040 Wien (Austria)}

\vspace{-5mm} 

\address{$^{c)}$ Theory Division, CERN \\
      CH - 1211 - Geneva 23 (Switzerland)}

\end{center}

\thispagestyle{empty}
\vspace{-10mm}

\abstract{Topological field theories of Schwarz-type generally admit 
symmetries whose algebra 
does not close off-shell, e.g. 
the basic symmetries of BF models or vector supersymmetry of 
the gauge-fixed action for Chern-Simons theory
(this symmetry being at the origin of the 
perturbative finiteness of the theory).
We present a detailed discussion of
all these symmetries within 
the algebraic approach to the Batalin-Vilkovisky 
formalism.
Moreover, we discuss the general algebraic 
construction of topological models of both Schwarz- and Witten-type.}

\end{titlepage}
\addtocounter{footnote}{-3}

\tableofcontents

\newpage 

\setcounter{page}{1}

\section{Introduction} 
 
The present paper is devoted to the 
 algebraic construction \cite{Ikemori:1992qz} 
 and to the symmetries
 of topological field theories of Schwarz-type 
 (see \cite{brt} for a review of the latter theories). 
The classical action of some of these models 
(e.g. BF models in a space-time of 
dimension $d\geq 4$ \cite{bfm, MyersPeriwal}) 
admits on-shell reducible symmetries and thus leads 
to a BRST-operator which is only nilpotent on-shell. 
Moreover, 
the gauge-fixed action for these models
admits a supersymmetry-like invariance, 
the so-called vector supersymmetry (VSUSY), which 
also generates an on-shell algebra 
\cite{Delduc:1989ft}-\cite{PiguetSorella}.
Such on-shell invariances raise problems upon quantization
of these theories. 

The Lagrangian {\em Batalin-Vilkovisky (BV) -formalism} 
represents a systematic approach to this problematics \cite{bv}. 
(For a short summary, see reference \cite{bbow}.)
In fact, in this canonical (symplectic) setting,  
all fields are supplemented from the beginning on with {\em antifields}
and these additional variables ensure off-shell 
closure of the symmetry algebras.  
The antifields of the BV-formalism correspond to the 
external sources of the standard 
BRST-approach and can be expressed in terms of 
the latter. In this way,  
one recovers symmetry algebras for the basic fields 
whose closure is guaranteed by the external sources.  
The latter transformations coincide with those obtained 
by the action of the {\em linearized Slavnov-Taylor operator}
in the standard BRST-approach. 

As matter of fact \cite{gms, brt}, topological field theories 
of Schwarz-type provide a
neat application for the BV-formalism which is often discussed in quite
general  terms in the literature \cite{sw}. 
The algebraic approach to this formalism was pioneered by 
H.~Ikemori \cite{Ikemori:1992qz} and applied to various 
models in the sequel \cite{Baulieu:1996ep, bau98} 
(see also \cite{wallet} for earlier work and \cite{carval} 
for an interesting field-theoretic interpretation). 
We will incorporate VSUSY in this framework 
and show that this yields a major simplification with respect to 
the algebraic approach to the 
BRST-formalism \cite{Gieres:2000pv}, thus simplifying 
the study of the renormalization and finiteness properties  
of topological models \cite{Pisar:2000xs}.

In the present work, we restrict our attention to models 
in flat $d$-dimensional space-time, but a generalization of VSUSY 
to arbitrary manifolds can be achieved  \cite{cgp}. 
In fact, the latter allows to tackle the relationship 
between BF models and gravity \cite{bfm, witten}  where 
a VSUSY-like invariance also exists \cite{p}. 

The fundamental ingredients of the algebraic approach 
\cite{Ikemori:1992qz,Baulieu:1996ep}
are {\em extended forms} (corresponding to the {\em complete ladders} 
of the BV-formalism) : the latter can be used to write down 
action functionals as well as 
{\em horizontality conditions} or {\em Russian formulae}
\cite{Baulieu:1984ih, Bertlmann}
summarizing the basic symmetries of the action. 
The essential tool for describing  VSUSY-transformations
is given by the so-called {\em $\emptyset$-symmetry condition}  
introduced in reference \cite{Gieres:2000pv}. 

Our paper is organized as follows. 
In section 3, we discuss in detail the examples 
of $3$-dimensional Chern-Simons theory and 
of the $4$-dimensional BF model \cite{Ikemori:1992qz}.
In doing so, we will make 
contact at all stages with previous studies 
of these models within the BRST- or BV-setting. In particular, 
we will elaborate on VSUSY and carry out explicitly 
the elimination 
of antifields so as to allow for a comparison with the results
obtained from other lines of reasoning. 
To anticipate our conclusions, we already indicate that 
the algebraic approach allows to recover various known 
results (or slight generalizations thereof) almost effortless, 
in a systematic way and in a quite compact form. 
As a by-product, we will present a novel interpretation of the 
VSUSY-algebra in subsection 3.2.3.  
Our study of concrete examples provides the hindsight 
for formulating some general principles for the algebraic 
construction of topological models. This will be the subject 
of section 5 where we also summarize 
the different classes of topological models 
of both Schwarz- and Witten-type 
which can be obtained along these lines.

\section{General setting} 
   
The models to be discussed admit a Lie algebra of symmetries 
and the involved fields  
are $p$-forms with values in this Lie algebra.
In particular, we always have 
the Yang-Mills connection $1$-form $A$ and the associated curvature 
$2$-form $F = dA + {1 \over 2} [A, A]$. 
The field strength
of any additional 
field $\varphi$ is given by 
 its covariant derivative $D \varphi = d  \varphi + [A, \varphi ]$. 
We will only be concerned with the classical theory and the fields 
occurring in the initial invariant action of a model  
will be referred to as 
{\em classical} fields.

\section{Example 1: Chern-Simons theory in ${\bf R}^3$}

\subsection{Symmetries of the classical action} 

\subsubsection{The model and its symmetries}

The action 
\be
S_{\rm inv} [A]  = {1 \over 2} \intR{3} 
\tr \{ AdA + \frac{2}{3}AAA \}  
\label{CSinv}
\ee 
 is invariant under infinitesimal 
 gauge transformations, $\delta A = Dc$,  
 and it leads to 
the equation of motion $F=0$, i.e. a zero-curvature condition 
for the connection $A$. 
In expression (\ref{CSinv}) and in the following, 
the wedge product sign is omitted. 

The gauge invariance of the functional (\ref{CSinv}) represents an 
off-shell, irreducible symmetry and therefore 
the BV-description of this invariance 
leads, up to minor modifications, to the same 
results as the BRST-approach.
In fact, by starting our study with a symmetry of this simple type, 
 we can best recognize the precise correspondence between both 
formalisms.

\subsubsection{Geometric framework of BRST- and BV-approaches} 

By way of motivation and to fix the notation, 
we recall a few facts concerning the 
geometric framework of the {\em BRST-approach}
\cite{PiguetSorella, Bertlmann}. 
In this setting, infinitesimal symmetry parameters are turned
into ghost fields.  
Thus, the geometric sector of the Chern-Simons model involves the classical
field $A$ and the ghost field $c$ associated to infinitesimal gauge
transformations. 
Lower and upper indices of a field label its form degree
and ghost-number, respectively. 
For each field, the ghost-number is added to the form degree 
in order to define 
a {\em total degree} and all commutators are assumed to be graded 
with respect to this degree. 
The BRST-operator $s$ increases the ghost-number by one 
unit, but it does not modify the form degree of fields. 
In view of the definition of 
Green functions or the formulation of 
the Slavnov-Taylor identity, 
one also introduces external sources $\gamma_2^{-1}$ 
and $\sigma_3^{-2}$ which couple to the 
(non-linear) BRST-transformations 
of $A$ and $c$, respectively: this amounts to the addition of a term 
\[
S_{\rm ext} = 
\intR{3} \tr \{ \gamma_2^{-1} sA + \sigma_3^{-2} sc \}
\]
to the action. The latter contribution is $s$-invariant since 
the operator $s$ is 
nilpotent and since the external fields $\gamma$ and $\sigma$
are assumed to be $s$-inert. 

In the {\em BV-approach} that we will consider here, one starts
with the fields $(\Phi^a)=(A,c)$ together with the corresponding {\em
antifields} $(\Phi^*_a)=(A^* ,c^*) 
\equiv (A^{-1}_{2} , A^{-2}_{3})$
which have the {\em same index structure} as the sources 
$(\gamma_2^{-1}, \sigma_3^{-2})$. All of these fields then 
define  the geometric or {\em minimal sector} of the theory.
In the sequel (section 3.1.4), 
$s$-variations are defined for all of these 
variables, the transformations of $A^*$ and $c^*$ being non-trivial.  
Direct contact with the BRST-approach 
is established at a later stage 
(after carrying out the gauge-fixing procedure) 
by eliminating the
antifields $(A^*, c^*)$ in terms of external sources  $(\gamma, \sigma)$
- see sections 3.1.8 and 3.1.9.

\subsubsection{Generalized fields and derivatives} 

In the BRST-approach, the ghost field $c$ 
is added to the connection $A$ in order to define the generalized 
 field $\tilde A = A+c$. 
In the BV-approach, 
the corresponding antifields are also included 
\cite{Ikemori:1992qz}  
so as to obtain the {\em generalized field} (or {\em extended form}) 
\begin{eqnarray} 
        \tilde A \!\!\!&=&\!\!\! A^{-2}_{3} + A^{-1}_{2} + A + c
\nonumber  \\
\label{com}
 \!\!\!&=&\!\!\!  c^*+A^*+A+c	\, .
\end{eqnarray}
The latter contains fields of {\em all} form degrees which are
allowed by three-dimensional space-time (and therefore it also involves 
fields with {\em negative} ghost-number). It is referred to as a 
{\em complete ladder} and it can be viewed as a ``self-dual'' quantity
since it involves the basic fields $A$ and $c$ 
together with their antifields \cite{Ikemori:1992qz,Baulieu:1996ep},
 see section 5 below. 

The $s$-differential is added to the exterior derivative $d$ 
 so as to define 
the {\em generalized derivative} $\tilde d=d+s$.
In the same vein, one introduces the 
{\em generalized field strengths} of 
$\tilde A$ and $\varphi$:  
\bea
        \tilde F=\tilde d \tilde A + {1 \over 2} [ \tilde A, \tilde A ]
	\qquad , \qquad 
\tilde D\varphi=\tilde d\varphi+[\tilde A,\varphi]	
	\, .
\label{genF}
\eea
Actually, it also proves useful \cite{Baulieu:1996ep}
to define the generalized fields 
\be
   F^{\tilde A} = d\tilde A + {1 \over 2} [\tilde A , \tilde A ]
	\qquad , \qquad    
   D^{\tilde A}\varphi = d\varphi+[\tilde A,\varphi]
	\, , 
\label{genDF}
\ee
which satisfy the Bianchi identities 
  \bea 
 D^{\tilde A}  F^{\tilde A}  =0 
  \qquad , \qquad
D^{\tilde A}D^{\tilde A}  \varphi = [F^{\tilde A} ,\varphi ]
  \, .
  \label{bb}
\eea

\subsubsection{Derivation of $s$-variations 
   from a horizontality condition}

Just like the standard BRST-transformations, the 
{\em $s$-variations} in the BV-framework 
can be obtained from a 
{\em horizontality condition}. 
For the topological models under consideration, 
the latter is obtained by 
replacing ordinary fields by generalized fields in the 
equations of motion of the model \cite{Ikemori:1992qz}.
Thus, for Chern-Simons theory,  
 one imposes the generalized zero-curvature condition
\be
\label{ze}
\tilde F=0
\ .
\ee
By virtue of the definition of $\tilde F$, this relation is equivalent to 
\be
\label{sti}
s\tilde A = -F^{\tilde A }
\, ,
\ee
i.e.  condition (\ref{ze}) 
determines the $s$-variations of all component fields of $\tilde A$
\cite{Ikemori:1992qz,Baulieu:1996ep}.  
The resulting transformations read  
\be
  \ba{rclcrcl}
   s c^* \!\!\!&=&\!\!\! -DA^* - [c,c^*]
 &\quad ,\quad & s A        \!\!\!&=&\!\!\! -Dc 
  \\
            s A^* \!\!\!&=&\!\!\! -F - [c,A^* ]
 &\quad ,\quad &        s c   \!\!\!&=&\!\!\! - {1 \over 2} [c,c]
  \ea
\label{CSBRST}
\ee
and they are nilpotent by virtue of the 
Bianchi identities (\ref{bb}), see reference
\cite{Baulieu:1996ep, Gieres:2000pv}. 

In equations (\ref{CSBRST}), 
we recognize the standard BRST-transformations of 
the basic fields $A$ and $c$. 
If we were to eliminate the antifields by setting them to zero
together with their $s$-variations, 
relations (\ref{CSBRST}) would lead to the classical equations of motion of
the model.  Accordingly, this method of elimination is inappropriate
and another procedure will be considered after performing the gauge-fixing.

\subsubsection{Construction of the minimal BV-action}

Our goal is to  extend the invariant action (\ref{CSinv})
to a $s$-invariant functional 
for the fields $ (\Phi^a; \Phi^*_a) = (A,c\, ; A^*,c^*)$
of the minimal sector. 
For its construction, we
follow a purely algebraic reasoning,  
i.e. we ignore the space-time dimension. 
The Chern-Simons Lagrangian satisfies 
\[
d \; \tr \{ AdA + {2 \over 3} A^3 \} = \tr (FF)
\, ,  
\]
henceforth a relation of the same form holds for the 
`tilde'-variables : 
\[
\tilde d \; \tr \{ \tilde A \tilde d\tilde A 
+ {2 \over 3} \tilde A^3  \} 
= \tr (\tilde F\tilde F) 
\, .
\]
By virtue of the horizontality condition $\tilde F =0$, 
the right-hand-side vanishes and, as a consequence of $\tilde d = d+s$, 
we have  
\be
\label{cons}
s \; \tr \{ \tilde A \tilde d \tilde A 
+ {2 \over 3} \tilde A ^3 \} 
= 
-d \; \tr \{ \tilde A \tilde d \tilde A 
+ {2 \over 3}  \tilde A^3 \}
\, .
\ee
The polynomial $\tr \{...\}$ involves $\tilde d \tilde A$ and therefore
it {\em explicitly} contains $s$-variations of fields.
Since we do not want such terms to appear in our action, we will 
eliminate them in terms of others.  
By virtue of $\tilde d = d+s$, 
the left-hand-side of equation (\ref{cons}) 
reads 
\[
\mbox{LHS} = 
s \; \tr \{ \tilde A d \tilde A 
+ {2 \over 3} \tilde A ^3 \}
+ I 
\]
with 
\[
I \equiv 
s \; \tr \{ \tilde A s \tilde A \} 
= \tr \{ s \tilde A \, s \tilde A \} 
= \tr \{ F^{\tilde A} \, F^{\tilde A} \} 
= d \; \tr \{ \tilde A d \tilde A 
+ {2 \over 3} \tilde A^3 \}
\, .
\]
Using $0= \tilde F = \tilde d \tilde A + \tilde A^2$, the right-hand-side 
of equation (\ref{cons}) can be rewritten as 
\[
\mbox{RHS} = 
d \; \tr \{ {1 \over 3} \tilde A^3  \} 
\, . 
\]
Thus, as a final result, we find the {\em cocycle condition} 
\be
\label{coco}
s \; \tr \{ \tilde A d \tilde A 
+ {2 \over 3} \tilde A ^3 \}
= 
-d \; \tr \{ \tilde A d \tilde A 
+ {1 \over 3}  \tilde A^3 \} 
\, .
\ee
The  polynomial $\tr \{...\}$ 
on the LHS involves contributions of different 
form degrees. 
 After integrating its $3$-form part over space-time \cite{Ikemori:1992qz},  
we obtain the so-called {\em minimal BV-action} which is $s$-invariant
by virtue of relation (\ref{coco}):   
\bea
S_{\rm min} [\Phi^a ; \Phi^*_a ] 
\!\!\!& \equiv &\!\!\! {1 \over 2} \intR{3} \left.
\tr \{ \tilde{A}d\tilde{A} + \frac{2}{3}\tilde A \tilde A
\tilde A  \} \right|^{0}_{3} \nm \\
\!\!\!&=&\!\!\! S_{\rm inv} [A] + 
\intR{3} \tr 
\{ A^* Dc + c^* cc \}
\label{CSactionmin}
\\
\!\!\!&=&\!\!\! S_{\rm inv} [A]  
- \intR{3} \tr 
\{ \Phi^*_a \, s\Phi^a \}
\, .
\nm 
\eea
The $s$-invariance of the second term is non-trivial, since 
the antifields $A^*$ and $c^*$ transform non-trivially under the $s$-operation.  
The simple structure of this term 
reflects the fact that the basic fields 
$(\Phi^a) =(A,c)$ transform among themselves: if they were to mix with the 
antifields, additional terms would appear in (\ref{CSactionmin}),  
see the BF model below.

\subsubsection{Gauge-fixing} 

We fix the gauge by imposing the Lorentz condition $d\star \! A =0$ 
where $\star A$ denotes the Hodge-dual of the $1$-form $A$. 
This condition 
is implemented in the action by adding 
the $s$-exact functional $S_{\rm gf} = s \Psi_{\rm gf}$ where the 
{\em gauge-fixing fermion} $\Psi_{\rm gf}$ is given (in the Landau gauge) by 
\be
\label{land}
        \Psi_{\rm gf}=\intR{3}\tr\left\{\bar c \left(d
	\star \! A\right)\right\}
	\, .
\ee
The antighost $\bar c$ appearing in this expression forms a 
BRST-doublet with an auxiliary field $b$, i.e. 
\be 
\label{cb}
s\bar c = b 
\qquad , \qquad 
sb =0
\, .
\ee
As for the geometric sector, one also introduces 
the corresponding antifields $(\bar c^{\, *}, \, b^*)$ and couples  
them to the $s$-variations of $\bar c$ and $b$, respectively:  
\bea
\label{au}
S_{\rm aux} [ \bar c , b \, ; \bar c^{\, *} , b^* ]
\!\!\!& \equiv &\!\!\! -\intR{3}\tr \, \{ \bar c^{\, *} s \bar c +
b^* sb \}
\\
\!\!\!&=&\!\!\! -\intR{3}\tr \, \{ \bar c^{\, *}b \}
\, .
\nm
\eea
The antifields $(\bar c^{\, *}, b^*)$ are again assumed to form a 
BRST-doublet, but one transforming ``the other way round",  
\be 
\label{bc}
s\bar c^{\, *} = 0
\qquad , \qquad 
sb^* = \bar c^{\, *}
\, .
\ee
This guarantees the $s$-invariance of $S_{\rm aux}$.

By adding the functional (\ref{au}) to the minimal action
(\ref{CSactionmin}), 
one obtains the 
so-called {\em non-minimal BV-action}
which depends on the fields 
$(\Phi^A) = (A,c,\bar c,b)$ and the associated antifields
$(\Phi_A^*)$ : 
\bea
S_{\rm nm} [ \Phi^A \, ; \Phi_A^* ]
\!\!\!&\equiv&\!\!\! 
S_{\rm min} + S_{\rm aux}
\nonumber 
\\
\!\!\!&=&\!\!\! 
S_{\rm inv}
 +\intR{3}\tr \{ A^*Dc + c^*cc - \bar c^{\, *} b \}
\nonumber
\\
\!\!\!&=&\!\!\! 
S_{\rm inv}
  - \intR{3} \tr \{ \Phi_A ^* s\Phi^A  \}
\, . 
\label{CSactionnm}
\eea
Note that this action is $s$-invariant and that it does not include 
the gauge-fixing functional $S_{\rm gf}= s \Psi _{\rm gf}$.

\subsubsection{BV-interpretation} 

Let 
$(\Phi^A)=(\Phi^a,\bar C^\alpha, \Pi^\alpha)$ 
collectively denote all fields, i.e. the classical and ghost fields
$(\Phi^a)$ defining the minimal sector,  
the antighosts $\bar C^\alpha$ and the multiplier fields $\Pi^\alpha$.
Accordingly, let 
$(\Phi^*_A)=(\Phi^*_a,\bar C^*_\alpha,\Pi^*_\alpha)$ denote the associated
antifields and let $(\Theta ^A) = (\Phi^A ; \Phi^*_A)$. 
Quite generally, if $\Phi^A$ has index structure 
$(\Phi^A)_p^g$, then the corresponding antifield has 
index structure $(\Phi^*_A)_{d-p}^{-g-1}$
where $d$ denotes the space-time dimension. 
Accordingly, for a space-time  ${\cal M}_d$ 
of odd (even) dimension,  
the fields and their antifields have the same (opposite) parity. 

For any two functionals $X[ \Theta ^A]$ and $Y[ \Theta ^A]$ of the 
fields $(\Theta ^A)$,
the {\em BV-bracket} is the graded bracket defined by 
\bea
\left\{X,Y\right\}
=
\int_{{\cal M}_d} 
 \tr \left\{
\W{X}{\Phi^A}\W{Y}{\Phi^*_A}\pm\W{Y}{\Phi^A}\W{X}{\Phi^*_A}  \right\}
\, ,
\label{braket} 
\eea 
where the sign depends on the Grassmann parity of $X$ and 
$Y$. (Our convention to use {\em left} functional derivatives
differs from the one which is frequently used in the literature \cite{sw}.)

Let $\Gamma[\Phi^A,\Phi^*_A]$ be the non-minimal BV-action (\ref{CSactionnm})
as defined on ${\cal M}_d = {\bf R}^3$. 
Then, the
latter is a solution of the 
{\em classical BV master equation} 
\bea
\{\Gamma,\Gamma \}=0 
\qquad {\rm i.e.} \quad \int_{{\cal M}_d} 
 \tr \left\{ \W{\Gamma }{\Phi^A} \,  \W{\Gamma }{\Phi_A^*} 
 \right\} =0  
\label{master} 
\eea
and the {\em $s$-variations of fields and antifields} are given by 
\bea 
s\Theta^A= \{\Gamma,\Theta^A \}
\, ,  
\label{BRS} 
\eea 
i.e.  
\be
s \Phi^A =  - \W{\Gamma }{\Phi_A^*} 
\qquad , \qquad 
s \Phi_A^* =  - \W{\Gamma }{\Phi^A} 
\label{svar}
  \ee
for $(\Phi^A ) = ( A, c , \bar c , b)$.
Indeed, the explicit expressions following from (\ref{svar}) 
with $\Gamma = S_{\rm nm}$ coincide 
with the transformation laws (\ref{CSBRST}), 
(\ref{cb}) and (\ref{bc}).  
Since $s\Gamma =  \{\Gamma,\Gamma \}$, 
the master equation (\ref{master}) expresses the $s$-invariance 
of $\Gamma$. The off-shell nilpotency of the $s$-operator 
can be viewed as a 
consequence of the graded Jacobi identity for the BV-bracket.

As a matter of fact, the functional 
$S_{\rm min}$ already represents a solution
of the master equation 
which only depends on the variables 
$(\Phi ^a ; \Phi_a^*) = 
(A, c \, ; A^*, c^*)$, i.e. 
\be
s \Phi^a = - \W{S_{\rm min} }{\Phi_a^*} 
\qquad , \qquad 
s \Phi_a^* =  - \W{S_{\rm min} }{\Phi^a} 
\, .
\label{sv}
  \ee
The latter result confirms the identification between antifields
and forms of negative ghost-number considered in equation (\ref{com}).
It also explains why the choice $\Phi _a ^* = 0 = s\Phi _a ^*$
implies the classical 
equations of motion \cite{Baulieu:1996ep}.

\subsubsection{Elimination of antifields}

Since antifields have been associated to all fields,  
{\em external sources} are also introduced 
for each field (and not only for those transforming 
non-linearly under the $s$-operation, as is usually done in the
BRST-approach). The sources associated to 
$(\Phi^A) = (A,c,\bar c,b)$ are denoted by 
$(\rho_A) \equiv (\gamma,\sigma,\bar \sigma,\lambda)$. 
Then, the antifields
$(\Phi_A^*) = (A^*,c^*,\bar c^{\, *},b^*)$ are eliminated in terms of 
these sources by virtue of the {\em general prescription}  
\bea
\Phi^*_A 
\, = \, - \hat \rho_A 
\, \equiv \, 
- \rho_A \, + \, 
(-1)^{(d+1)|\Phi^A|+d} 
\ \W{\Psi_{\rm gf}}{\Phi^A}
\, , 
\label{exa} 
\eea 
where $|\Phi^A|$ denotes the total degree of the field $\Phi ^A$
and $d$ the space-time dimension (i.e. $d=3$ in our example). 
Thus, we get 
\be
\label{axe}
\ba{lcl}
A^* = -( \gamma+ \star  \, d\bar c) \ \equiv \  -\hat \gamma 
\, &\quad , \quad &
c^* = -\sigma
  \\
\bar c^{\, *} = -\left(\bar \sigma + d \star \! A\right) \ \equiv \ 
-\hat {\bar \sigma}  
&\quad , \quad & 
        b^* = -\lambda
	\, ,
\ea
  \ee
i.e. the antifields become the ``hatted" sources 
$(\hat{\rho} _A ) \equiv 
(\hat{\gamma} , \sigma , \hat{\bar{\sigma}} , \lambda)$
which amount to a repara\-metrization of certain sources. 

After substituting these expressions into the non-minimal BV-action
(\ref{CSactionnm}), one 
obtains the following functional which only depends on the ``hatted"
sources: 
\bea
\Sigma  \!\!\!&=&\!\!\!
S_{\rm inv} 
+ \intR{3}\tr \{ 
\hat{\rho} _A s \Phi^A \}
\nm \\
\!\!\!&=&\!\!\!
S_{\rm inv} 
+ \intR{3}\tr \{ s \Phi_A \W{\Psi_{\rm gf}}{\Phi^A} \} 
+ \intR{3}\tr \{ \rho _A s \Phi^A \}
\nm \\ 
\!\!\!&=&\!\!\!
S_{\rm inv} 
+ \intR{3}\tr \{ bd \star \!  A + \bar{c}d \star \!  Dc \}
  + \intR{3}\tr \{  \gamma sA + \sigma sc + \bar\sigma s\bar c +\lambda sb  \}
\, .
\label{CSactionG}
\eea
Actually, the transformation law $sb=0$ implies that 
the last term vanishes so that $\Sigma$
does not depend on the source $\lambda$.
Thus, we have  
\bea
        \Sigma[\Phi^A,\rho_A] \equiv  
  \left.S_{\rm nm}\right|_{\Phi^*_A}
	=S_{\rm inv}+S_{\rm gf}+S_{\rm ext}
	\, ,
\label{gamma}
\eea
where $S_{\rm inv}$ represents the classical, gauge invariant action, 
$S_{\rm gf} =s\Psi_{\rm gf}$  the associated 
gauge-fixing functional and $S_{\rm ext}$ the 
linear coupling of 
external sources $\rho_A$ to the $s$-variations of the fields
$\Phi^A$.
The action $\Sigma$ represents the {\em $s$-invariant vertex functional} 
(at the classical level) and coincides with the result 
obtained from the 
usual BRST procedure (e.g. see \cite{PiguetSorella}),  
except for the presence of the 
external sources $(\bar \sigma,\lambda)$ coupling to the $s$-variations 
 of the BRST-doublet 
$(\bar c , b)$.

By substituting expressions 
(\ref{axe}) into the $s$-variations of the antifields,
we obtain those of the sources: the variables 
$(\lambda , \hat{\bar\sigma})$ transform like a BRST-doublet, 
\be
\label{dob}
s\lambda=\hat{\bar\sigma}
\qquad , \qquad s\hat{\bar\sigma}=0
\, , 
\ee
and thereby  the 
{\em $s$-variations of the sources} take the explicit form  
\be
  \ba{lcllcllcl}    
s \gamma \!\!\!&=&\!\!\!  F-[c,\hat\gamma]- \star  \, db 
\quad &,& \quad
s \bar \sigma \!\!\!&=&\!\!\! d \star  \! Dc
\\
s \sigma \!\!\!&=&\!\!\! -D\hat\gamma-[c,\sigma] 
\quad &,& \quad
s \lambda \!\!\!&=&\!\!\! \bar \sigma +d\star \! A
\, .
\ea
  \ee
The transformation laws of all variables can be summarized by 
\be
\label{sro}
s \Phi^A =  \W{\Sigma }{\rho_A} 
\qquad , \qquad 
s \rho_A =  \W{\Sigma }{\Phi^A} 
  \ee
with $(\Phi^A ) = ( A, c , \bar c , b)$ and 
$(\rho_A ) = ( \gamma, \sigma , \bar{\sigma} , \lambda)$. 
These relations are the relict of the BV-variations (\ref{svar})
after the elimination of all antifields. 

Obviously, the $s$-variations (\ref{sro}) determine the {\em linearized
Slavnov-Taylor operator},  i.e. 
\bea 
s = \SLT_\Sigma \equiv \int_{{\bf R}^3}
\tr\left\{\LST{\Phi^A}{\rho_A}\right\}
\, .
\label{LST} 
\eea 
 Thus, the BV master equation (\ref{master}) becomes the 
{\em Slavnov-Taylor identity} 
\be
\label{mst}
\SLT(\Sigma) \equiv \int_{{\bf R}^3}
\tr\left\{\ST{\Phi^A}{\rho_A}\right\}=0
\, , 
\ee 
which ensures that $s^2=0$ off-shell and that 
$s\Sigma=\SLT_\Sigma(\Sigma)=2\SLT(\Sigma)=0$.
This interpretation of the master equation is a cornerstone 
of the theory and is further discussed in the literature, both at the
classical and quantum  level \cite{bv, sw}.  
Here, we only put forward two points. First, we note that  
the hatted sources associated to antighosts and multipliers
form BRST-doublets (see eqs.(\ref{dob})) which simplifies the cohomogical
analysis of the quantum theory. Second, we remark that the 
 Slavnov-Taylor identity (\ref{mst}) of the BV-approach 
 has a more symmetric form than the one  
 of the BRST-approach where one only introduces
 external sources for those fields which transform non-linearly, 
 i.e. $A$ and $c$.

\subsubsection{BV versus BRST}
 
 Let us summarize the conclusions that we can draw from 
 our discussion of the Chern-Simons theory 
 at the classical level (and which are in accordance with 
 the general results \cite{bv, sw}). 
For an off-shell, irreducible symmetry (like YM-invariance),  
the differences between the BV-approach and
the BRST-procedure are two-fold : 
\begin{itemize}
   \item The $s$-operator of the BV-formalism represents 
   the linearized Slavnov-Taylor operator: unlike the standard 
   BRST-differential, this operator  
   does {\em not} leave the external sources invariant. 
   (The non-trivial transformation laws of the sources follow
   from the non-trivial $s$-variations of antifields given  
   by equations (\ref{svar}), after the elimination of antifields
   in terms of sources by virtue of the gauge fermion $\Psi_{\rm gf}$.) 
     \item In the BV-approach, one introduces sources for all fields, 
   not only for those transforming non-linearly, as one usually  
   does in the BRST-framework. 
Of course, the latter framework also 
{\em allows} for the inclusion of such sources: although 
they are not particularly useful, they lead to a more symmetric
expression for the  Slavnov-Taylor identity (and also for 
the Ward identities, see equation (\ref{qui}) below 
and comments thereafter).
\end{itemize}

\subsection{VSUSY}

\subsubsection{VSUSY-transformations}
   
After performing the gauge-fixing for the invariant action 
(\ref{CSinv})  
in a Landau-type gauge, 
the gauge-fixed action is  invariant under VSUSY-transformations
\cite{Delduc:1989ft}.
In this section, we will introduce these transformations
as well as their algebra and,   
in the next section, we will discuss the induced variation 
of the BV-action. 

At the infinitesimal level, VSUSY-transformations are 
described by an operator $\delta _{\tau}$  
where $\tau \equiv \tau^\mu \partial_{\mu}$ 
 represents a constant $s$-invariant vector field of ghost-number
zero. The variation $\delta_{\tau}$ 
acts as an antiderivation which lowers
the ghost-number by one unit and which anticommutes with $d$.
The operators $s$ and $\delta _{\tau}$  satisfy a   
graded algebra of Wess-Zumino type :   
\begin{equation}
\label{algebra}
[ s, \delta_{\tau} ] \, = \, {\cal L} _{\tau} \, + \, 
\mbox{equations of motion} \ .
\end{equation}
Here, $\Li = [ i_{\tau}, d]$ 
denotes the Lie derivative with respect to 
the vector field $\tau$ and 
$i_{\tau}$ the interior product with $\tau$.

In the {\em BRST-approach},
the VSUSY-variations for topological models of Schwarz-type 
can be derived from the so-called {\em $\emptyset$-type symmetry 
  condition} \cite{Gieres:2000pv}
\be 
\label{vsusy}
\delta_{\tau} \tilde A = i_{\tau}  \tilde A
\ .
\ee 
In the {\em BV-approach}, we start from the same expression,   
the only difference being that 
$\tilde A$ now involves both fields of positive and 
negative ghost-number.  
Substitution of the expansion (\ref{com}) into (\ref{vsusy}) yields 
the VSUSY-variations in the geometric sector,  
\be
\label{dst}
\del c = \id A 
 \quad , \quad 
 \del A = \id A^*    
   \quad , \quad 
   \del A^* = \id c^*   
  \quad , \quad 
  \del c^* = 0
  \, . 
  \ee
We note that, if two fields are related by VSUSY 
(e.g.  $c \stackrel{\del \, }{\to} A$), then the 
corresponding antifields are related ``the other way round''
(i.e. $A^* \stackrel{\del \, }{\to} c^*$).
This feature represents a useful {\em guideline}
for dealing with more complex models or field contents. 

Using (\ref{vsusy}) and $s\tilde A = - F^{\tilde A}$, it can be 
explicitly shown that $[s, \del ] \tilde A = \Li \tilde A$, i.e. 
the VSUSY-algebra is satisfied off-shell for all fields
of the geometric sector. 
(In fact, this result holds by construction \cite{Gieres:2000pv}.) 

We now turn to the transformation laws 
of the remaining fields and antifields. 
The $\delta_{\tau}$-variation of $\bar c$ has to 
vanish for dimensional and ghost-number reasons 
(``there is nothing it can transform into''). 
If we require the VSUSY-algebra to be satisfied off-shell for all fields, 
we readily obtain the variation of $b$ : 
\be
\label{dbl}
\del \bar c = 0 \qquad \Rightarrow \qquad 
\del b = \del s \bar c = (\Li - s \del ) \bar c = \Li \bar c
\ .
\ee
As for the $s$-variations, 
the associated doublet of antifields $( \bar c^{\, *}, b^*)$ is assumed to 
transform ``the other way round" (in accordance with the general 
guideline indicated above) : 
\be
\label{dud}
\del b^* = 0 \qquad , \qquad 
\del \bar c^{\, *} = \Li b^*
\ .
\ee
  
After eliminating all antifields in terms of sources 
by virtue of equations (\ref{axe}),  
 we get  the  {\em VSUSY-variations}  
\be
\label{csba}
   \ba{rclcrcl}
        \del c \!\!\!&=&\!\!\!  \id A 
&\quad ,\quad& 
 \del \hat{\gamma} \!\!\!&=&\!\!\!   \del \gamma = \id \sigma
  \\
  \del A \!\!\!&=&\!\!\! -\id \hat\gamma = -\id (\gamma + \star d\bar c) 
  &\quad ,\quad & 
 \del \sigma \!\!\!&=&\!\!\! 0
\ea
  \ee
and 
 \be
\label{bis}
\ba{lcl}
\del \bar c = 0  
&\quad ,\quad & 
\del b= \Li\bar c
\\
\del \bar \sigma = \Li \lambda  + d\star  \id \hat \gamma 
&  \quad ,\quad & 
  \del \lambda =0
\, ,
\ea
\ee
where the variation of $\bar \sigma$ is a consequence of 
$\del \hat{\bar \sigma}=\Li\lambda$. 
It can be checked explicitly that the VSUSY-algebra holds 
off-shell for all fields and sources.

If we set the sources to zero, we recover the results of the 
standard BRST-approach for the VSUSY-variations of  the 
fields $(A, c , \bar c , b)$ \cite{Delduc:1989ft}.
For these fields,  the $s$-variations 
of the BV- and BRST-approaches coincide so that we also recover the 
on-shell VSUSY-algebra.
If sources are included in the BRST-framework for the discussion
of Ward identities, a different argumentation from the one 
considered here leads to the same VSUSY-variations for
$(A, c , \bar c , b)$ and $(\gamma, \sigma)$ 
\cite{delluc,PiguetSorella}.
In fact, external sources (antifields) somehow 
play the same r\^ole as {\em auxiliary fields} in supersymmetric field
theories in that they lead to a symmetry algebra which closes off-shell 
\cite{gms, bbow}.

\subsubsection{Ward identity}

The transformation law (\ref{vsusy}) induces the following variation of the 
minimal BV-action (\ref{CSactionmin}): 
\begin{eqnarray*} 
\del S_{\rm min}  \!\!\!&=&\!\!\! 
{1 \over 2} \int_{{\bf R}^3} \tr 
\left. \{ \tilde A \Li \tilde A \} \right| _3^{-1} 
\\ 
 \!\!\!&=&\!\!\! 
 \int_{{\bf R}^3} \tr 
 \{ A^* \Li A + c^* \Li c  \} 
 \ .
\end{eqnarray*} 
The transformations (\ref{dbl}),(\ref{dud}) yield a similar expression
for the variation of the auxiliary action (\ref{au}): 
\[
\del S_{\rm aux} = 
 \int_{{\bf R}^3} \tr 
 \{ b^* \Li b + \bar c^{\, *} \Li \bar c \}
\ . 
\]
Thus, the non-minimal BV-action $S_{\rm nm} = S_{\rm min} + S_{\rm aux}$
satisfies a 
{\em broken Ward identity:} 
\be 
\label{bwid}
\WI_\tau  S_{\rm nm}
 \equiv \int_{{\bf R}^3} \tr \left\{\del \Phi^A
\W{S_{\rm nm} }{\Phi^A}+\del \Phi^*_A \W{S_{\rm nm}}{\Phi^*_A}\right\}
= 
\int_{{\bf R}^3} \tr \left\{\Phi^*_A\Li\Phi^A \right\}
\, . 
\ee
The breaking is linear in the fields (and also in the antifields).
Henceforth, it is unproblematic for the quantum theory 
since `insertions' that are linear in quantum fields are not
renormalized by quantum corrections. 
The result (\ref{bwid}) can also be derived 
from expression (\ref{CSactionnm})  
 by substituting the variations (\ref{dst})-(\ref{dud})  
and using  $[ s, \del ] = \Li$.

After elimination of the antifields, the Ward identity (\ref{bwid})
takes the form 
\be
 \label{Ward}
 \WI_\tau \Sigma
 \equiv 
  \int_{{\bf R}^3} \tr \left\{\del \Phi^A \W{\Sigma}{\Phi^A}
 +\del \rho_A \W{\Sigma}{\rho_A}\right\}
 = \Delta_{\tau}  
\ee
with 
\be
\label{qui}
\Delta_{\tau} = 
 - \int_{{\bf R}^3} \tr \left\{ \rho_A  \Li\Phi^A\right\}
 = 
 - \intR{3} \tr \left\{\gamma\Li A
+ \sigma \Li c + \lambda \Li b 
+ \bar\sigma\Li\bar c\right\}  
 \, . 
\ee
The Ward identity (\ref{Ward}), with $\del \Phi^A$ and $\del \rho_A$
given by eqs. (\ref{csba})(\ref{bis}), 
has the same form as the one found in the BRST-framework
\cite{PiguetSorella, vilar} where the sources $\lambda, \bar \sigma$ are not
considered. 
Yet, it is their inclusion which leads to the quite symmetric
expression (\ref{qui}). 

The fact that the  breaking is linear in the quantum fields
and in the sources only holds in the {\em Landau gauge} 
that we have chosen in the gauge-fixing fermion (\ref{land}): 
in a different gauge, implemented by the gauge-fermion
\[
\Psi_{\rm gf} = \int_{{\bf R}^3} \tr \{ \bar c
(d \star A - {\alpha \over 2} \star b) \} 
\qquad {\rm with} \ \; \alpha \in {\bf R}^* 
\, , 
\]
 the breaking term $\Delta_{\tau}$ is non-linear
 in the quantum fields. 

To summarize the two previous sections, we can say that 
the BV-approach readily leads to a VSUSY-algebra which closes 
off-shell and to a Ward identity which is broken by a term that 
is linear in the quantum fields and sources (in the Landau gauge).

\subsubsection{On the algebra of symmetries} 

Before proceeding further, we come back shortly to the algebra of 
symmetries which may be summarized as follows. 
The basic operators 
\[
d \equiv \delta_1^0 \quad , \quad s \equiv \delta_0^1
\quad , \quad
i_{\tau} \equiv \delta_{-1}^0 \quad , \quad \del \equiv \delta_0^{-1} 
\]
modify the form degree $p$ and ghost-number $g$ of a field 
$\varphi$ according to 
\[
\delta_m^n \left( \varphi_p^g \right) =  \left(\delta  \varphi
\right)_{p+m}^{g+n} 
\]
and satisfy the graded algebra 
\[
[ \delta_m^n , \delta_k^l ] \, = \, 
\delta_{m+k, 0} \, \delta_{n+l,0} \, \Li
\ .  
\]

It is interesting
to compare the action of the operators $\del$ and 
$s$ on the basic fields. For this purpose, we decompose $s$ 
according to 
$s= s_0+s_1$ where $s_0$ and $s_1$ represent, respectively, 
the linear and non-linear parts of the operator. 
By virtue of   $s\tilde A = - F^{\tilde A}$,
the action of $s_0$ on the generalized field $\tilde A$ 
is given by 
\be
\label{hc0}
s_0 \tilde A = - d \tilde A 
\, .
\ee
Comparison with (\ref{vsusy}) shows that each of the operators 
$s_0$ and $\del$ 
acts in the same fashion on all fields occurring in the  
 expansion $\tilde A$. 
 However, the two operators act into opposite directions
 inside the ladder $\tilde A$ : 
while $s_0$ increases the ghost-number by one unit, 
$\del$ lowers it by the same amount, 
\be
\label{ncd}
\tilde A \ = \ 
\begin{array}{c}
 \stackrel{s_0}{\longrightarrow} \\    
 c^* + A^* + A + c  \\
\stackrel{\longleftarrow}{\del}  
\end{array} 
\ , 
\ee
both operators 
being related by 
\begin{equation} 
\label{vsu}
[s_0, \del ] = \Li
 \ .
 \end{equation}

The linear part  $s_0$ of the $s$-operator (which already 
determines the non-abelian structure of the 
theory to a large extent \cite{abel})
also allows for a unified formulation of all symmetries. 
To present this  geometric description, we define
(in analogy to $\tilde d = d +s$)
\begin{eqnarray}
   \tilde d_0 \!\!\!& \equiv &\!\!\! d + s_0 
   \nonumber \\
  \tilde i_{\tau} \!\!\!& \equiv &\!\!\! i_{\tau}  - \del
\ . 
\end{eqnarray}
The horizontality condition (\ref{hc0}) defining $s_0$ and the symmetry
condition (\ref{vsusy}) defining $\del$ are then equivalent to 
\be
 \tilde d_0 \tilde A = 0 \qquad , \qquad  
 \tilde i_{\tau} \tilde A =0
\ee
and the compatibility condition for these two equations, 
\[
0= [  \tilde d_0 , \tilde i_{\tau} ]  
= [ d, i_{\tau} ] - [ s_0 , \del ]
\]
is the VSUSY-algebra relation (\ref{vsu}).

 \subsubsection{BV versus BRST} 

Quite generally, we can say the following.
Once external sources (associated to non-linear field variations)
are introduced in the {\em BRST-framework} for discussing 
Ward identities, one recovers the same results for
VSUSY-transformations as in the BV-approach
and also the same type of expression 
for the breaking of VSUSY. Yet, in the 
{\em BV-framework} where sources are introduced for all fields
under the disguise of antifields, the 
VSUSY-breaking term has a more symmetric 
form. 

Concerning the derivation of VSUSY-variations (for 
topological models of Schwarz-type) by virtue of 
the symmetry condition (\ref{vsusy}), the conclusion is as follows. 
This symmetry condition 
can be taken as a starting point in the standard BRST-formalism 
\cite{Gieres:2000pv}, but the derivation
of symmetry transformations is 
already involved for a simple model like Chern-Simons theory
due to the fact that one has to refer to its equations of motion.  
By contrast, the {\em BV-formalism} allows for a quite
simple and straightforward study of VSUSY.

\section{Example 2: BF model in ${\bf R}^4$}
   
The approach to the BF model closely follows the lines of the 
Chern-Simons theory, henceforth we will only emphasize the new features
that it exhibits. 

\subsection{Symmetries of the classical action}
   
\subsubsection{The model and its symmetries}

The BF model in  ${\bf R}^4$ 
 involves two gauge potentials: the YM $1$-form $A$ and 
the $2$-form potential $B\equiv B_2^0$, i.e. 
a Lie algebra-valued $2$-form transforming
under the adjoint representation of the gauge group.   
  The model is characterized by the action 
\be
  \label{fct}
  S_{\rm inv} [A,B] = \int_{\br} \tr \{BF\}
 \, , 
\ee
  which leads to the equations of motion 
\be
  \label{eqm}
F=0  
\qquad {\rm and} \qquad 
D B=0
\, .  
\ee  

The functional (\ref{fct}) 
is not only invariant under ordinary gauge transformations,
  but also under the local symmetry 
\be
  \label{es}
  \delta B = DB_1
  \, . 
  \ee
  By virtue of the second Bianchi identity 
  $D(DB_0) = [F, B_0]$ and the equation of motion 
$F=0$, the right-hand-side of (\ref{es}) is 
on-shell invariant under the transformation 
$\delta B_1 = D B _0 $. Thus, the 
symmetry (\ref{es}) is {\em one-stage reducible on-shell}.

\subsubsection{Horizontality conditions and $s$-transformations}

Apart from the ghost $c$ parametrizing ordinary gauge transformations, 
  we now have ghosts $B_1^1\equiv B_1$ and $B _0^2 \equiv B_0$ 
  parametrizing the reducible symmetry 
  (\ref{es}). Thus, one introduces 
 {\em generalized forms} \cite{Ikemori:1992qz} 
\be
\label{BFgenAB}
\begin{array}{rclcrcl}
\tilde A \!\!\!&=&\!\!\! A_4^{-3}+A_3^{-2}+A_2^{-1}+A+c 
&\quad , \quad & 
\tilde B \!\!\!&=&\!\!\! B_4^{-2}+B_3^{-1}+B+B_1^1+B^2_0
\\
\!\!\!&=&\!\!\! B_0 ^* +B_1 ^*+B^*+A+c
&\quad , \quad & 
\!\!\!&=&\!\!\! c^*+A^*+B+B_1 + B_0
\, , 
\end{array}
\ee
where $B_0^* \equiv (B_0)^*, \, B_1^* \equiv (B_1)^*$ and 
  where the identification of {\em antifields} has been 
performed as for the Chern-Simons theory, i.e. by 
considering  the index structure of all fields (see section 3.1.2). 
The gauge potentials 
$A$ and $B$ and, more generally, the generalized fields 
$\tilde A$ and $\tilde B$ 
can be viewed as dual to each other 
(see references \cite{Baulieu:1996ep,Ikemori:1992qz} and
section 5 below). 

In view of the equations of motion (\ref{eqm}), one postulates 
the  {\em horizontality conditions} \cite{Ikemori:1992qz}
\be
\tilde F=0  
\qquad {\rm and} \qquad 
\tilde D\tilde B=0
\, .  
\ee
These relations are equivalent to 
\be
  \label{sum}
s\tilde A=-F^{\tilde A}
\qquad {\rm and} \qquad 
s\tilde B=-D^{\tilde A}\tilde B
\ee
and thereby determine all $s$-variations : by substitution of 
expressions (\ref{BFgenAB}), one obtains 
\be
\ba{rclcrcl}
        sc \!\!\!&=&\!\!\! -{1 \over 2} [c,c]
  &\quad , \quad &
  \quad sB^* \!\!\!&=&\!\!\! -F-[c,B^*]
\\
        sA \!\!\!&=&\!\!\! -Dc 
  \ &\quad , \quad &
        sB_1^* \!\!\!&=&\!\!\! -DB^* -[c,B_1^* ]
\label{sb}
\\
 \!\!\!&&\!\!\! 
&\quad  \, \quad & 
   sB_0^* \!\!\!&=&\!\!\! -DB_1^*-[c,B_0^* ]-\bh[ B^*, B^* ]
\ea
\label{BFBRSTA} 
\ee
and
\bea
        sB_0  \!\!\!&=&\!\!\! -[c,B_0]
	\nm \\
        sB_1  \!\!\!&=&\!\!\! -DB_0-[c,B_1]
	\nm \\
        sB   \!\!\!&=&\!\!\! -DB_1-[c,B]-[B^* ,B_0] 
\label{BFBRSTB} \\
        sA^*  \!\!\!&=&\!\!\!
	-DB-[c,A^*]-[B^*,B_1]-[B_1^* ,B_0] 
	\nm \\
        sc^* \!\!\!&=&\!\!\!
	-DA^* -[c,c^*]-[B^* ,B]
	-[B_1^* ,B_1]-[B_0^* ,B_0]
	\, . \nm
\eea
The fields and antifields of the minimal
sector can be collected in $(\Phi^a)=(A,c,B,B_1,B_0)$ 
 and $(\Phi^*_a)=(A^*,c^*,B^*, B_1^*,B_0^*)$.
By construction, 
  the $s$-variations of these variables as given  by 
(\ref{sb}) and (\ref{BFBRSTB}) are  nilpotent off-shell.
  The fact that the transformation law 
  of the classical field $B$ involves the antifield $B^*$
  reflects the fact that the 
  symmetry algebra generated by (\ref{es}) 
  closes only on-shell.
  If all antifields are set to zero, one recovers the 
  standard BRST-transformations of  $(A,c,B,B_1,B_0)$
  which are only nilpotent on the mass-shell.

\subsubsection{Minimal BV-action}

Proceeding along the lines of section 3.1.5, we can 
extend the classical action (\ref{fct}). 
From the horizontality conditions (\ref{sum}), 
we obtain the {\em cocycle condition} \cite{Baulieu:1996ep}
\[
 s\; \tr \{ \tilde B F^{\tilde A} \} = - 
 d \; \tr \{ \tilde B F^{\tilde A} \}
\, , 
\]
which yields the $s$-invariant {\em minimal BV-action} 
\cite{Ikemori:1992qz,Baulieu:1996ep}
\be
S_{\rm min} [\Phi^a ; \Phi^*_a ] \equiv 
\int_{\br} \left. 
\tr  \{ \tilde B F^{\tilde A} \}
\right| _4^0 
\, .
\label{BFactionmin} 
\ee
Substitution of the expansions (\ref{BFgenAB}) leads to the 
explicit expression 
\be
\label{39}
S_{\rm min} = S_{\rm inv}
  - \int_{\br} \tr \{ \Phi^*_a \, s \Phi ^a \} 
- \bh  \int_{\br} \tr \{  B^* [ B^* ,B_0] \} 
\, , 
\ee
where the 
 last term reflects the antifield dependence of the transformation
 law of $B$. 
  We note that all of the $s$-variations 
(\ref{BFBRSTA}) and (\ref{BFBRSTB}) 
have the form BV-form (\ref{sv})
 which confirms the identification of antifields
made in (\ref{BFgenAB}).

\subsubsection{Gauge-fixing and elimination of antifields}

\paragraph{Gauge fermion and auxiliary fields}

Gauge-fixing of all symmetries, i.e. of YM-invariance and of the reducible 
symmetry of the $2$-form potential $B$,  requires 
a gauge fermion of the form 
\be
\Psi_{\rm gf} = \int_{\br}
\tr\left\{\bar c \, d\star \! A+\bar
c_1^{\, -1}d\star \! B+\bar c^{\, -2}d \star \! B_1 
+ \bar c^{\, 0} \left( 
d \star  \bar c_1^{\, -1} + \alpha \star  \pi^{-1} \right) 
\right\}
\quad (\alpha \in {\bf R}) 
\, . 
\ee
The involved 
antighosts $(\bar C^{\alpha}) \equiv \left( \bar c ,\bar c_1^{\, -1}, 
\bar c^{\, -2}, \bar c^{\, 0} \right)$
are supplemented with auxiliary fields 
$(\Pi^{\alpha} ) \equiv \left( b , \pi_1, \pi^{-1}, \pi^1 \right)$
so as to define BRST-doublets :
\be
\ba{rclcrclcrclcrcl}
        s\bar c \!\!\!&=&\!\!\! b
	& \quad , \quad &
s\bar c_1^{\, -1} \!\!\!&=&\!\!\! \pi_1
	& \quad , \quad & 	  
  s\bar c^{\, -2} \!\!\!&=&\!\!\!  \pi^{-1}
& \quad , \quad & 
s\bar c^{\, 0} \!\!\!&=&\!\!\! \pi^1  
  \\
	sb \!\!\!&=&\!\!\! 0
	& \quad , \quad &
               s\pi_1 \!\!\!&=&\!\!\! 0
	& \quad , \quad &
s\pi^{-1} \!\!\!&=&\!\!\! 0
&\quad , \quad &
s \pi^1 \!\!\!&=&\!\!\! 0
\, . 
\ea
\ee  
The corresponding  antifields transform in a dual way,  
\be
\ba{rclcrclcrclcrcl}
        s\bar c^{\, *} \!\!\!&=&\!\!\! 0
	& \quad , \quad &
s(\bar c_1^{\, -1})^* \!\!\!&=&\!\!\! 0
	& \quad , \quad & 	  
  s(\bar c^{\, -2})^* \!\!\!&=&\!\!\! 0
& \quad , \quad & 
s(\bar c^{\, 0})^* \!\!\!&=&\!\!\! 0
  \\
	sb^* \!\!\!&=&\!\!\! \bar c^{\, *}
	& \quad , \quad &
               s(\pi_1)^* \!\!\!&=&\!\!\! -(\bar c_1^{\, -1})^*
	& \quad , \quad &
s(\pi^{-1})^* \!\!\!&=&\!\!\! -(\bar c^{\, -2})^*
&\quad , \quad &
s (\pi^1)^* \!\!\!&=&\!\!\! -(\bar c^{\, 0})^*
\ea
\ee
and thereby ensure the $s$-invariance of the functional 
\bea 
S_{\rm aux}=
- \int_{\br} \tr \{ (\bar C^{\alpha})^* \Pi _{\alpha} \} 
\, , 
\label{BFactionaux} 
\eea
 which gives rise to the non-minimal action 
$S_{\rm nm}=S_{\rm min}+S_{\rm aux}$.

\paragraph{Elimination of antifields} 

Altogether, we have the fields  
$(\Phi^A) = (\Phi^a, \bar C^{\alpha} , \Pi^{\alpha})$ 
with 
\be
(\Phi^a) =  \left( A,c,B,B_1,B_0 \right) 
 \quad , \quad 
(\bar C^{\alpha}) = \left( \bar c ,\bar c_1^{\, -1}, 
			  \bar c^{\, -2}, \bar c^{\, 0} \right)
\quad , \quad 
(\Pi^{\alpha} ) = \left( b , \pi_1, \pi^{-1}, \pi^1 \right)
\ee
and the associated 
external sources $(\rho_A )$ are to be denoted as follows :  
\be
\label{sour}
\left( \gamma,\sigma,\rho_2^{-1},\rho_3^{-2}, \rho_4^{-3} \right) 
\qquad , \qquad 
\left( \bar\sigma ,\bar\sigma_3^0 ,\bar\sigma_4^1,\bar\sigma_4^{-1} \right)
 \qquad , \qquad 
\left( \lambda , \lambda_3^{-1} , \lambda_4^0 ,\lambda_4^{-2} \right) 
\, .
\ee 
The antifields $(\Phi^*_A)$ will 
now be expressed in terms of these  sources by virtue 
of the prescription (\ref{exa}) with $d=4$.  
For the antifields of the minimal sector, this entails 
\be
\ba{rclcrcl} 
A^*  \!\!\!&=&\!\!\! - ( \gamma - \star \, d\bar c) \, \equiv \, -\hat\gamma
&\ \quad , \quad &
B^*  \!\!\!&=&\!\!\! -(\rho_2^{-1} + \star \, d\bar c_1^{\, -1}) \, \equiv
\, -\hat\rho_2^{\, -1}
\\
\label{46}
c^*  \!\!\!&=&\!\!\! -\sigma
&\quad , \quad &	
 B_1^*    \!\!\!&=&\!\!\! -(\rho_3^{-2}- \star  \, d\bar c^{\, -2})
 \, \equiv \, -\hat\rho_3^{\, -2}
  \\
  & &\!\!\! 
&\quad \, \quad & 
 B_0^*   \!\!\!   &=& \!\!\! -\rho_4^{-3}
 \, ,
\ea 
\ee
whereas the antifields associated to antighost and multiplier fields are
given by 
\be
\ba{rclcrcl}
\bar c^{\, *} \!\!\!&=&\!\!\! -(\bar\sigma+d \star \! A)\, \equiv \,
-\hat{\bar\sigma} &\quad , \quad & 
 b^* \!\!\!&=&\!\!\! -\lambda 
 \nm \\
 (\bar c_1^{\, -1})^* \!\!\!&=&\!\!\! - (\bar\sigma_3^0 - d \star  \! B - \star 
 \, d\bar c^{\, 0} ) \, \equiv \, -\hat{\bar\sigma}_3^0 
&\quad , \quad &   
(\pi_1)^* \!\!\!&=&\!\!\! -\lambda_3^{-1} 
\nm \\
(\bar c^{\, -2})^*  \!\!\!&=&\!\!\! -(\bar\sigma_4^1-d \star \! B_1) 
  \, \equiv \, -\hat{\bar\sigma}_4^1 
& \quad , \quad &     
  (\pi^{-1})^* \!\!\!&=&\!\!\! -(\lambda_4^0+\alpha \star \! \bar c^{\, 0}) 
 \, \equiv \, -\hat\lambda_4^0 
 \label{47}
 \\
( \bar c ^{\, 0} ) ^* \!\!\!&=&\!\!\!  - ( \bar{\sigma} _4^{ -1} 
			      - \alpha \star \pi^{-1}
-d \star \bar c_1^{\, -1}) \, \equiv \, -\hat{\bar\sigma}_4^{-1}
&\quad , \quad & 
  (\pi^1)^* \!\!\!&=&\!\!\! -\lambda_4^{-2}
	\, .
\ea
\ee

\paragraph{Vertex functional}

The gauge-fixed action including external sources is obtained 
  from $S_{\rm nm} = S_{\rm min} + S_{\rm aux}$
by eliminating  antifields
according to relations (\ref{46})(\ref{47}). This leads to 
\bea
 \Sigma \!\!\!&=&\!\!\!
S_{\rm inv}
+ \int_{\br} \tr  \{\hat\rho_A s\Phi^A \}+S_{\rm mod}
 \nm\\
\!\!\!&=&\!\!\!
  S_{\rm inv} +s\Psi_{\rm gf}
+ \int_{\br} \tr \{\rho_A s\Phi^A \} + S_{\rm mod}
 \nm\\
\!\!\!&=&\!\!\!
  S_{\rm inv} +   S_{\rm gf}  + S_{\rm ext} + S_{\rm mod}
\, ,
\label{mods}
\eea
where 
\bea
S_{\rm mod}=-\bh\int_{\br} \tr \{ B_0[\hat\rho_2^{-1},\hat\rho_2^{-1}] \} 
\eea
is related to the fact
that the $s$-variation of $B$ 
exhibits an antifield dependence, see equation (\ref{BFBRSTB}).  
For $\rho_A =0$, expression (\ref{mods}) coincides with the one of reference
\cite{gms}
in which external sources are introduced at a different stage.

\paragraph{$s$-variations}

After elimination of all antifields, 
the $s$-variations of the basic fields 
$(A,c,B, B_1, B_0)$ are exactly the same as before
except for the fact that 
$sB$ now depends on a (hatted) source rather than an antifield: 
\be
sB = -DB_1 - [c,B] + [ \hat\rho_2^{-1} , B_0 ]
\, .
\ee
The sources associated to the basic fields transform as
\bea
s\gamma
\!\!\!&=&\!\!\!
DB-[\hat\rho_3^{-2},B_0]-[\hat\rho_2^{-1},B_1]
-[c,\hat\gamma]- \star \, db
\nm\\
s\sigma
\!\!\!&=&\!\!\!
-D\hat\gamma-[\rho_4^{-3},B_0]-[\hat\rho_3^{-2},B_1]
-[\hat\rho_2^{-1},B]-[c,\sigma]
\nm\\
s\rho_2^{-1} \!\!\!&=&\!\!\! F-[c,\hat\rho_2^{-1}]+ \star \, d\pi_1
\nm\\
s\rho_3^{-2} \!\!\!&=&\!\!\! -D\hat\rho_2^{-1}-[c,\hat\rho_3^{-2}]- \star \, d\pi^{-1}
\nm\\
s\rho_4^{-3} \!\!\!&=&\!\!\! -D\hat\rho_3^{-2}-[c,\rho_4^{-3}]
+\bh [\hat\rho_2^{-1},\hat\rho_2^{-1}]
\eea
and those associated to the antighosts and multipliers 
transform as 
\be
 \ba{rclcrcl}
s \bar\sigma \!\!\!&=&\!\!\! -d \star \! Dc
&\quad , \quad & 
s \lambda \!\!\!&=&\!\!\! \bar \sigma+d \star \! A 
\\
s \bar\sigma_3^0 \!\!\!&=&\!\!\! d\star \! \left( DB_1+[c,B]
-[\hat\rho_2^{-1},B_0] \right) -\star \, d\pi^1 
&\quad , \quad & 
s \lambda_3^{-1} \!\!\!&=&\!\!\! 
-\bar\sigma_3^0+d\star \! B+\star \,  d\bar c^{\, 0} 
\\
 s \bar\sigma_4^1 \!\!\!&=&\!\!\! d\star \! \left( DB_0+[c,B_1^1] \right)
&\quad , \quad &  
s \lambda_4^0 \!\!\!&=&\!\!\! -\bar \sigma_4^1+d\star \! B_1^1-\alpha \star \pi^1 
\\
s \bar \sigma _4^{-1} \!\!\!&=&\!\!\! -d\star \pi_1     
&\quad , \quad & 
s \lambda_4^{-2} \!\!\!&=&\!\!\! -\bar\sigma_4^{-1}+\alpha \star \! \pi^{-1}
+d\star  \bar c_1^{\, -1}.
\ea
\ee
The $s$-variations of fields and sources all have the form (\ref{sro}),
henceforth 
the $s$-operator may 
again be identified with the 
  linearized Slavnov-Taylor operator $\SLT_\Sigma$.

\paragraph{BV versus BRST} 
   
   The general conclusions drawn from the Chern-Simons theory also hold
   in the present case. 
   An extra feature of the BV-formulation for the 
   BF model (which involves 
   a reducible symmetry) is that the    
  $s$-variation of the classical field $B$ depends on sources (thereby
  ensuring the {\em off-shell nilpotency} of the $s$-operator). 
 Another facet of this issue 
   is the presence of the functional $S_{\rm mod}$ 
    in the vertex functional.
    While the BV-approach automatically produces such contributions
    which are non-linear in the external sources, they have to be added
    ``by hand'' in the standard BRST-framework,  
   e.g. see \cite{Delduc:1989ft, gms, becchi}. 
   
   It should be noted that an off-shell formulation 
   for the basic $s$-variations can eventually be given within the 
   BRST-framework by mimicking the BV-approach, see reference  
\cite{PiguetSorella}.

\subsection{VSUSY}

\paragraph{VSUSY-transformations of fields}

As in the standard BRST-approach \cite{Gieres:2000pv}, 
we start from the 
$\emptyset$-symmetry conditions 
\be
\label{bfe}
\del\tilde A=\id \tilde A
\qquad , \qquad 
\del \tilde B=\id \tilde B
\, .
\ee
After spelling out these relations in terms of component fields 
and eliminating the antifields
in terms of sources, we obtain the {\em VSUSY-variations of the basic
fields},  
\be
\label{sbf1}
\ba{rclcrcl} 
        \del c \!\!\!&=&\!\!\! \id A
	&\quad , \quad & 
 \del B_0 \!\!\!&=&\!\!\! \id B_1
  \\
        \del A \!\!\!&=&\!\!\! -\id\hat\rho_2^{-1}
&\quad , \quad & 
        \del B_1 \!\!\!&=&\!\!\! \id B
\\
 \!\!\!&&\!\!\! 	&\quad \, \quad & \!\!\!
        \del B \!\!\!&=&\!\!\!  -\id\hat\gamma
\ . 
\ea
\ee
and the {\em variations of the associated sources} :  
\be
\label{sbf2}
\ba{rclcrcl}
        \del \hat \gamma \!\!\!&=&\!\!\!  
	 \id \sigma
	&\quad , \quad & 
	        \del \sigma\!\!\!&=&\!\!\!0
		\\
        \del \hat \rho^{-1}_2 \!\!\!&=&\!\!\! \id \hat \rho^{-2}_3
	&\quad , \quad & 
        \del \hat \rho^{-2}_3 \!\!\!&=&\!\!\! \id \rho^{-3}_4
\label{gui}
\\
 \!\!\!&&\!\!\! &\quad \, \quad & 
        \del \rho^{-3}_4 \!\!\!&=&\!\!\! 0
	\, .
\ea
\ee

Taking the results of Chern-Simons theory as a guideline
(see eq.(\ref{csba})), 
we now define the VSUSY-variations of the antighosts
$(\bar C ^{\alpha}) 
= \left( \bar c, \bar c_1^{\, -1}, \bar c^{\, -2},\bar c^{\, 0} \right)$
in such a way that relations (\ref{gui}) hold for the unhatted sources,
i.e. such that we have  
\be
\ba{rclcrcl}
        \del  \gamma \!\!\!&=&\!\!\!  
	 \id \sigma
	&\quad , \quad & 
	        \del \sigma\!\!\!&=&\!\!\!0
		\\
\label{wie}
\del \rho^{-1}_2 \!\!\!&=&\!\!\! \id  \rho^{-2}_3
	&\quad , \quad & 
        \del \rho^{-2}_3 \!\!\!&=&\!\!\! \id \rho^{-3}_4
\\
 \!\!\!&&\!\!\! &\quad \, \quad & 
        \del \rho^{-3}_4 \!\!\!&=&\!\!\! 0
	\, .
\ea
\ee
The transformation laws of the multipliers $(\Pi ^{\alpha})$ 
are then determined by the requirement that 
the VSUSY-algebra is satisfied, see equation (\ref{dbl}).
Altogether, we find the following 
{\em variations of antighosts and multipliers} :  
\be
\label{vam}
\ba{rclcrcl}
  \del \bar c \!\!\!&=&\!\!\!0
  	&\quad , \quad & 
	\del b \!\!\!&=&\!\!\! \Li\bar c
  \\
\del \bar c_1^{\, -1} \!\!\!&=&\!\!\! \g \bar c^{\, -2}
	&\quad , \quad & 
\del \pi_1  
\!\!\!&=&\!\!\! \Li\bar c_1^{\, -1}+\g\pi^{-1}
\\
        \del \bar c^{\, -2} \!\!\!&=&\!\!\! 0
		&\quad , \quad & 
		\del \pi^{-1} \!\!\!&=&\!\!\! \Li\bar c^{\, -2}
	\\
        \del \bar c^{\, 0} \!\!\!&=&\!\!\! 0
		&\quad , \quad & 
		\del \pi^1 \!\!\!&=&\!\!\! \Li \bar c^{\, 0}
	\, .
\ea
\ee
Here, $\g=  \tau^{\mu} g_{\mu \nu} dx^{\nu}$ denotes 
 the $1$-form associated to the vector field $\tau$ 
by virtue of a space-time metric $(g_{\mu \nu})$ 
\cite{PiguetSorella, Gieres:2000pv}.

If we set all sources to zero, we recover the transformation laws 
and on-shell VSUSY-algebra of the standard BRST-approach
\cite{Gieres:2000pv}.  If sources are included in the latter 
framework for the discussion of Ward identities, 
considerations different from ours lead to the introduction of
the hatted sources (\ref{46}) and to the variations 
(\ref{sbf1})-(\ref{vam}) \cite{gms}.

To conclude, we come to 
the {\em $\del$-variations of the sources (\ref{sour}) 
associated to the doublet fields} $(\bar C ^{\alpha}, \Pi ^{\alpha})$.  
According to the general guideline indicated after 
equation (\ref{dst}), these sources (antifields) are assumed 
to transform ``the other way round", in the opposite direction 
as the fields, see eq.(\ref{dud}):
\be
\label{last}
\ba{rclcrcl} 
\del \bar\sigma \!\!\!&=&\!\!\! \Li\lambda-d \star \! \id\hat\rho_2^{-1}
	&\quad , \quad & 
\del\lambda \!\!\!&=&\!\!\! 0
\\
\del \bar\sigma_3^0 \!\!\!&=&\!\!\! -\Li\lambda_3^{-1}+d\star \! \id\hat\gamma
	&\quad , \quad & 
\del \lambda_3^{-1} \!\!\!&=&\!\!\! 0 
\\
\del \bar\sigma_4^1 \!\!\!&=&\!\!\! -\Li\lambda_4^0-\g\bar\sigma_3^0
+(1-\alpha)\star \! \Li \bar c^{\, 0}
	&\quad , \quad & 
	\del \lambda_4^0 \!\!\!&=&\!\!\! - \g \lambda_3^{-1} 
\\
\del \bar\sigma_4^{-1} \!\!\!&=&\!\!\! 
-\Li \lambda_4^{-2}-d\star \! \g\bar c^{\, -2}
+\alpha \star \! \Li\bar c^{\, -2}
	&\quad , \quad & 
	\del \lambda_4^{-2} \!\!\!&=&\!\!\! 0
\, .
\ea
\ee
By construction, the VSUSY-algebra is satisfied off-shell for all 
fields and sources.

\paragraph{Ward identity}

The $\del$-variations of fields and antifields induce 
the {\em broken Ward identity}  
\be 
\WI_\tau S_{\rm nm} 
= \int_{\br}  \tr \{ \pm \Phi^*_A \Li\Phi^A \}
  \, ,
\ee
which takes the following 
form after elimination of antifields: 
\be 
\WI_\tau \Sigma
= \int_{\br} \tr \{ (-1)^{|\rho_A|} \, \rho_A\Li\Phi^A \}
  \, . 
\ee
Thus, the final result has the same form as for 
Chern-Simons theory, i.e. we have a breaking which is linear 
in the sources and in the quantum fields  \cite{gms, vilar}. 
It is worthwhile to note that this result has been obtained 
for an arbitrary value
 of the gauge parameter $\alpha$. This is in contrast to the usual 
formulation where VSUSY puts some restrictions on the gauge parameter
\cite{gms, PiguetSorella, vilar}.

\section{General case} 

The topological field theories studied in the previous sections
allowed us to make a detailed comparison between 
the BRST- and BV-approaches to the different types of symmetries
that are essential for discussing perturbative renormalization. 
They also 
provide concrete illustrations for an algebraic construction of 
topological models.
The goal of the present section is three-fold. 
First of all, to present some general principles summarizing 
the algebraic formalism considered so far (section 5.1.2). 
Second, to investigate which other models can be constructed  
using this approach and to determine their characteristic 
features (sections 5.1.3 and 5.1.4). 
Finally, we wish to provide some general expressions 
and strategies applicable to all models under consideration 
(sections 5.1.6 and 5.2).

\subsection{Symmetries of the classical action}

\subsubsection{Generalized forms and duality}

Since the algebraic construction
involves a gauge field and, more generally, $p$-form potentials, 
we first recall the general framework 
for $d$-dimensional space-time  ${\cal M}_d$ presented
in reference \cite{Baulieu:1996ep}. 

Let us consider $p \in \{ 0,1,...,d \}$. 
A $p$-form gauge potential $X_p \equiv X_p^0$
with values in a Lie algebra 
gives rise to a 
{\em generalized form} 
  \be
\tilde X_p=\sum_{q=0}^d
X_{d-q}^{p-d+q}=X_d^{p-d}+X_{d-1}^{p-d+1}+\ldots+X_p+\ldots+X^p_0
\, ,
\label{genX} 
\ee
 which involves all ghosts and ``ghosts for ghosts" as well 
as some fields with negative ghost-number. However, 
in general, the 
 index structure of the latter fields does not allow us to identify 
 them with the antifields associated to the fields 
  appearing in $\tilde X_p$. 
 Rather one has to introduce a so-called {\em dual form} $\tilde Y_{d-p-1}$ 
    with an analogous expansion,  
\be 
\tilde Y_{d-p-1}=\sum_{q=0}^d
Y_{d-q}^{-p-1+q}=Y_d^{-p-1}+\ldots+Y_{d-p-1}+\ldots+Y^{d-p-1}_0
\, . 
\label{genY} 
\ee 
The generalized forms (\ref{genX}) and (\ref{genY}) 
are dual to each
other in the sense that the fields with negative ghost-number  
contained in the first one 
  are the antifields associated to 
  the fields with positive ghost-number contained in the second one 
and vice versa, i.e. 
\bea 
\left(X_{p-q}^q\right)^* \!\!\!&=&\!\!\! Y_{d-(p-q)}^{-q-1}
\; \qquad \ {\rm for} 
\quad q=0,\ldots,p
\nm \\
\left(Y_{d-p-1-q}^q\right)^* \!\!\!&=&\!\!\! X_{p+1+q}^{-q-1} 
\qquad \quad {\rm for} \quad q=0,\ldots,d-p-1
\, .
\eea
For instance, 
for  $d=4$, the ladder $\tilde A$ is dual to the ladder
$\tilde B$ while $\tilde A$ is self-dual for $d=3$.

\subsubsection{Field content and construction of models}
   
   All of the models to be considered 
   involve a gauge field $A$, hence a generalized form 
   $\tilde A \equiv \tilde X_1$ and the dual form 
$\tilde B \equiv \tilde Y_{d-2}$. 
Eventually, additional dual pairs $(\tilde X_p, \tilde Y_{d-p-1})$ with 
$0 \leq p \leq d-1$ 
can be included and coupled to the 
generalized gauge field $\tilde A$. 

We now summarize the general algebraic 
procedure for constructing models 
\cite{Ikemori:1992qz,Baulieu:1996ep}. 
\begin{enumerate}
   \item One imposes {\em horizontality conditions} on $\tilde A, \tilde B, 
\tilde X, \tilde Y$, i.e. conditions on their field strengths 
$\tilde F, \tilde D \tilde B, \tilde D \tilde X, \tilde D \tilde Y$
which are compatible with the Bianchi identities
\[
\tilde D \tilde F =0 
\qquad {\rm and} \qquad 
\tilde D ^2 \tilde \Omega = [ \tilde F, \tilde
\Omega ]
\quad {\rm for} \quad 
\tilde \Omega = \tilde B, \tilde X, \tilde Y
\ . 
\]
   This determines nilpotent $s$-variations for the components 
   of $\tilde A, \tilde B, \tilde X, \tilde Y$.
In practice, the horizontality conditions are nothing but the 
{\em tilted equations of motion} of the model 
to be defined in the next step: thus, the horizontality conditions 
fix both the symmetries and the dynamics. 
\item 
One looks for a {\em generalized Lagrangian density}
$\tilde {\cal L}$, i.e. a generalized $p$-form which depends on 
$\tilde A, \tilde B, ...$ and their exterior derivatives
$d \tilde A, d \tilde B, ...$
and which satisfies the {\em cocycle condition} 
$s \tilde {\cal L}= d(..)$ where $s$ denotes the operator 
defined in the first step. 
Then 
\be 
\label{poly}
S_{\rm min} [\Phi^a, \Phi_a^*] \equiv \int_{{\cal M}_d}  
\left. \tilde {\cal L} 
\right| _d^0  
\ee
represents an {\em $s$-invariant action} extending the {\em classical
action} \be
\label{polinv}
S_{\rm inv} [\Phi^a]
\equiv S_{\rm min} [\Phi^a, \Phi_a^*=0]
\ . 
\ee
Moreover, the $s$-variations defined in the first step coincide
with those generated by the functional 
(\ref{poly}) according to relations 
(\ref{sv}), i.e. $S_{\rm min}$ solves the BV master 
equation\footnote{The explanation of this fact represents an intriguing
question.}. 
\end{enumerate} 
Since the $s$-operator is defined in terms of conditions involving 
the covariant quantities 
$\tilde F, \tilde D \tilde B, ...$, the polynomial 
$\tilde {\cal L}$ 
depends on $d \tilde A, d\tilde B, ...$ by virtue of the field strengths 
$F^{\tilde A}, D^{\tilde A} \tilde B ,...$. 
By construction, the classical action (\ref{polinv}) is invariant under the 
{\em standard BRST-transformations} 
$s_0 \Phi^a \equiv \left. \left( s\Phi^a \right) \right| _{\Phi^*_a =0}$.  
If $s\Phi^a$ involves $\Phi^*_a$, then $s_0$ is only 
nilpotent on the mass-shell.  

As emphasized in reference \cite{Baulieu:1996ep}, the algebraic 
approach proceeds in the opposite order than the usual 
{\em BV-algorithm}. In fact, the latter starts with a classical action 
$S_{\rm inv} [A,B,..]$ 
that is invariant under $s_0$-variations 
(which are, in general, only nilpotent on the mass-shell) 
and the goal then consists of explicitly 
determining an action 
\[
\Gamma [ \Phi^A; \Phi^*_A ] = S_{\rm inv} [A,B,..]
+ \sum_{q=1}^{n} \Phi^*_{A_1} \cdots \Phi^*_{A_q}
\Delta_q [\Phi^A] 
\, , 
\]
satisfying the master equation and generating nilpotent $s$-variations
by virtue of the definitions (\ref{svar}).

\subsubsection{Examples}

Mostly following references \cite{Ikemori:1992qz,Baulieu:1996ep}, 
we will now present an overview of models which can be 
constructed by the procedure outlined above.
As pointed out by L.~Baulieu \cite{Baulieu:1996ep, bau98}, 
this construction not only yields topological field 
theories of Schwarz-type, but also theories of Witten-type.
We will 
not spell out the explicit form of the cocycle condition for each model,
since the latter can easily be obtained from the given Lagrangian 
$\tilde {\cal L}$ by application of the $s$-operator. 
However, we note that (contrary to the indications 
in \cite{Baulieu:1996ep}), the cocycle condition does not always have 
the simple form $(s+d) \tilde {\cal L} =0$ as was already illustrated
by the Chern-Simons theory, see equation (\ref{coco}). 
The compatibility of horizontality conditions can readily
be verified for each example.

\paragraph{1) BF-model in $d\geq 2$}
       
This model involves the  pair 
$(\tilde A, \tilde B)$, the action being given by   
\be
\label{bfd}
S _{\rm min} \; = \; \int _{{\cal M}_d} 
\left. \tr  \{ \tilde B F^{\tilde A} \} \right| _d^0 
\; = \;  \int _{{\cal M}_d} 
\tr  \{ B F \}
\; + \; \mbox{s.t.} 
\ , 
\ee
where `s.t.' stands for source terms. 
The classical equations of motion are the zero-curvature conditions
$F=0=DB$.
For $d=2$, the field $B$ represents a $0$-form and does not have a 
local gauge symmetry (apart from ordinary gauge transformations)
      \cite{bfm, brt, gms}.

\paragraph{2) BF-model with cosmological constant}
For $d=3$ and $d= 4$, a term involving 
a real dimensionless parameter $\lambda$ can be added \cite{bfm} 
to the BF-action 
for the pair $(\tilde A, \tilde B)$. For ${\bf d=3}$, the 
minimal action reads \cite{Ikemori:1992qz} 
\be
   S _{\rm min} \; = \; \int _{{\cal M}_3} 
   \left. \tr  \{ \tilde B F^{\tilde A} 
   + {\lambda \over 3} \tilde B ^3 \}
\right| _3^0 
\; = \;  \int _{{\cal M}_3} 
\tr  \{ B F + {\lambda \over 3} B ^3 \}
\; + \; \mbox{s.t.} 
\ , 
\ee
which leads to the classical equations of motion 
$F+ \lambda B^2 = 0 = DB$. 
  
  For ${\bf d=4}$, the action 
   \be
   \label{tym}
   S _{\rm min} \; = \; \int _{{\cal M}_4} 
   \left. \tr  \{ \tilde B F^{\tilde A} 
   + {\lambda \over 2} \tilde B ^2 \}
\right| _4^0 
\; = \;  \int _{{\cal M}_4} 
\tr  \{ B F + {\lambda \over 2} B ^2 \}
\; + \; \mbox{s.t.} 
\ee   
leads to the complete equations of motion 
$F^{\tilde A} + \lambda \tilde B = 0 = \tilde D \tilde B$. 
From these, we can deduce, amongst others, that 
\be
\label{sbst}
sB^* = - [c, B^* ] - (F + \lambda B) 
\ . 
\ee
We now impose the complete equation of motion for the $2$-form potential
$B$, i.e. 
\be
\label{full} 
0 = \left. ( F^{\tilde A} + \lambda \tilde B) \right|_2^0 
= \ds{\delta S_{\rm min} \over \delta B} = - s B^* 
\ .
\ee
This implies that the field $B^*$ can be set to zero consistently.  
In addition, we choose $\lambda =-1$ for the sake of 
simplicity. From 
equations (\ref{full})(\ref{sbst}), we then conclude that 
$B=F$ and, by substitution into (\ref{tym}), we obtain 
   \be
\left. S _{\rm min} \right| _{{\delta S_{\rm min} \over \delta B} =0} 
    \; = \; \ds{1 \over 2} \int _{{\cal M}_4} 
   \tr  \{ FF \}
   \; + \; \sum_{\Phi^a = A,c,B_1,B_0} 
 \int _{{\cal M}_4} 
    \tr  \{ \Phi^*_a s  \Phi^a \}  
\ .
\ee   
This expression represents the minimal action 
associated to the topological invariant 
$\int _{{\cal M}_4} \tr  \{ FF \}$ 
whose gauge-fixing gives rise to Witten's 
{\em topological Yang-Mills theory} (TYM) \cite{brt}. 
  For the latter theory, both the BRST-algebra 
  \cite{bausing}
  and the  VSUSY-algebra \cite{Brandhuber:1994uf, Gieres:2000pv} 
  close off-shell 
  for different classes of gauge-fixings so that 
  the introduction of antifields 
    does not seem useful for studying the quantization 
  of the model. Yet, it is quite interesting that 
  TYM whose gauge-fixing procedure refers to self-{\em duality}
  conditions can be obtained from an action  
involving only a {\em dual} pair of potentials \cite{bau98}.

\paragraph{3a) BF-XY-model}
   
 For $d \geq 2$, one can add to the BF-model (\ref{bfd}) 
 some dual pairs 
 $(\tilde X_p, \tilde Y_{d-p-1})$ with $0 \leq p \leq d-1$ 
 coupling to
 $\tilde A $ according to \cite{Baulieu:1996ep}
 \be
   S_{\rm min} \; =\;  \int _{{\cal M}_d} 
   \left. \tr  \{ \tilde B F^{\tilde A} 
   + 
   \sum _{p=0}^{d-1} \tilde X_p D^{\tilde A} \tilde Y_{d-p-1} \}
\right| _d^0 
 \; =  \; \int_{{\cal M}_d} 
 \tr  \{ B F + \sum _{p=0}^{d-1} X_p D Y_{d-p-1} \}
 \, + \,  \mbox{s.t.} 
 \ . 
\ee
This action leads to the  classical equations of motion
$0=F=DB - \sum _{p=0}^{d-1} (-1)^p [X_p , Y_{d-p-1} ] = 
DX_p = DY_{d-p-1}$. It represents 
 a first order action that is 
   analogous to three-dimensional 
   Chern-Simons theory.

\paragraph{3b) BF-XY-model with BX-coupling}  
For any $d\geq 2$,   the $2$-form potential $X_2$
appearing in the previous model 
can  be coupled directly to $B$ 
  \cite{Baulieu:1996ep}
 with strength $\alpha \in {\bf R}$ : 
  \be
\label{mda}
  S_{\rm min} =  \int_{{\cal M}_d} 
   \left. \tr  \{ \tilde B (F^{\tilde A} + \alpha \tilde X_2)  
   + \tilde X_2 D^{\tilde A} \tilde Y_{d-3} \}
\right| _d^0 
 =   \int_{{\cal M}_d} 
\tr  \{ B (F + \alpha X_2) + X_2 D Y_{d-3}
   \}
+ \mbox{s.t.} 
\ .
\ee
  The classical equations of motion then take the form 
$0=F + \alpha X = DY + \alpha B = DB - [X,Y]$. 
By elimination of $X$ from the action functional 
by virtue of its algebraic equation of motion, 
one gets a classical action of the form 
   $ \int_{{\cal M} _d} d  \, \tr  \{ F Y_{d-3} \}$
which is analogous to TYM \cite{Baulieu:1996ep}. 
More specifically, for $d=3$, this action, 
$ \int_{{\cal M} _3} d  \, \tr  \{ F Y_{0} \}
=  \int_{{\cal M} _3} \tr  \{ F D Y_{0} \}$  
describes magnetic monopoles and its gauge-fixing via 
  Bogomolny's equations yields a topological 
  model that is closely related to four-dimensional TYM 
  \cite{bogomol}.

For ${\bf d=4}$, a ``dual'' form of the model (\ref{mda})
  is obtained  
by exchanging the generalized fields $\tilde B_2$ and $\tilde X_2$
in the pairs $(\tilde A_1, \tilde B_2), (\tilde X_2, \tilde Y_1)$: 
  \be
   S_{\rm min} =  \int_{{\cal M}_4} 
   \left. \tr  \{ \tilde X_2 (F^{\tilde A} + \alpha \tilde B_2)  
   + \tilde B_2 D^{\tilde A} \tilde Y_{1} \}
\right| _4^0 
\ .
\ee 
To this functional one can add a contribution 
$\int
\left. \tr {\cal F} (\tilde B_2 ) \right| _4^0$ of the form 
$s\int \Delta_4^{-1} + \int d(...)$. 
Different gauge-fixings then allow to recover the Lagrangian 
$\tr \{ \tau \, FF\}$ for TYM and the one of the dual theory 
defined by the duality transformation $\tau \to 1/ \tau$ 
(see reference \cite{bausha} for this and the following points).
The $\theta$-parameter of the theory can be 
adjusted by adding the topological invariant
\begin{eqnarray*} 
 S_{{\rm top}} & \!\!\! \equiv \!\!\!  & \int_{{\cal M}_4} 
 d\, \tr \{ a \, (AdA + \ds{2 \over 3} AAA ) \, +  \, 
b \, F Y_1 \, + \, c \, Y_1 DY_1 \} 
\\
&\!\!\! = \!\!\! &
a  \int_{{\cal M}_4} \tr \{ FF \} \, + \, 
b  \int_{{\cal M}_4} \tr \{ FDY_1 \} \, + \, 
c  \int_{{\cal M}_4} \tr \{ DY_1 \, DY_1 + F [Y_1, Y_1] \}
\end{eqnarray*} 
 with appropriately chosen complex parameters $a,b,c$.
 The different formulations of TYM in two and eight dimensions can be
 approached along the same lines.

\paragraph{3c) BF-XY-model with mixed couplings}   
 Several sets of pairs 
  $(\tilde X_p, \tilde Y_{d-p-1}),  (\tilde U_p, \tilde V_{d-p-1}),...$
 with $0 \leq p \leq d-1$ can be considered and coupled by terms 
 of the form $[X,Y], \, [X,U], \, B[X,U],...$ \cite{bau98}. 
For concreteness, we consider $d=6$ and independent pairs
  $(\tilde A, \tilde B_4),\,  
 (\tilde X_2, \tilde Y_3),  \, 
 (\tilde U_2, \tilde V_3)$, 
$(\tilde U^c_2, \tilde V^c_3)$ 
 with an action $S_{\rm min} = S_{\rm inv} + \mbox{s.t.}$ where  
  \be
   S_{\rm inv} =  \int_{{\cal M}_6} 
 \tr  \{ B (F + X) 
 + X ( D Y + [U,U^c]) + UDV - U^c DV^c +VV^c \}
\ .
\ee 
By substituting the algebraic equations of motion 
$0=F+X= V+DU^c=V^c+DU$ into the latter functional, we obtain
the following {\em six-dimensional topological model of Witten-type}
\cite{bau98}: 
  \bea
   S_{\rm inv} &=&  \int_{{\cal M}_6} 
 \tr  \{ DU \, DU^c - F [U,U^c] \} - 
  \int_{{\cal M}_6} \tr  \{  FDY \}
  \\
 &=&  \int_{{\cal M}_6} 
 d \, \tr  \{ U DU^c \} 
  -\int_{{\cal M}_6} d \, \tr  \{  FY \}
  \nonumber 
 \ .
\eea
The first part of this action admits 
BRST- and VSUSY-algebras which close off-shell and which have been studied
in references \cite{bau98, Gieres:2000dr}.

\paragraph{4) $3d$ Chern-Simons theory and extensions thereof}     
   For $d=3$, we can choose $\tilde X_1 = \tilde A = \tilde Y_1$ 
and consider the {\em Chern-Simons theory} as we did in section 3. 
 One can also combine this theory with the models considered above 
 \cite{Baulieu:1996ep}
or include 
  a term $\int_{{\cal M}_3} \tr  \{  X_1 D X_1 \}$
  \cite{MyersPeriwal}. 
The generalization of Chern-Simons theory
 to an arbitrary dimension \cite{MyersPeriwal} may be discussed as well 
using the algebraic approach \cite{Dayi:1993fk}.

\paragraph{5) Supersymmetric extensions of the previous models}  
 The algebraic formalism admits a supersymmetric extension 
 \cite{sasa} which should allow to discuss the supersymmetric 
 versions of the previous models, e.g. super BF models
 (see \cite{pasa} and references therein).

\subsubsection{General features}  

One may wonder what kind of field theoretic 
models can be constructed by the algebraic procedure summarized 
above and which generic features are shared by 
  all of the models that we listed.  
  Obviously, their {\em field content} 
is given by {\em $p$-form potentials} and they involve at least the 
connection $1$-form $A$. Only in three dimensions, 
the corresponding extended form $\tilde A$  
contains solely $A$, its ghost $c$ and the associated antifields 
$A^*, c^*$. Thus, $3$-manifolds are the only ones 
for which a model involving solely a Yang-Mills potential 
can be constructed. 
(Actually, such models can be obtained
indirectly in other dimensions by eliminating some fields, as
illustrated by example 2 above.) 

 Otherwise, a common feature of all models constructed above is that  
their {\em minimal action} (involving the classical fields $A,...$, 
the ghosts $c,...$  
 and the associated antifields $A^*,c^*,...$) can be written directly 
 in terms of {\em generalized fields} $\tilde A,...$ obeying some 
{\em generalized zero-curvature conditions}
 $0= \tilde F = ...$ \cite{vilar}.
This fact is related to the following one. 
The {\em dynamics} 
of fields is described by a  {\em metric-independent,  
first order action}, 
the kinetic term being of the form 
$AdA, BdA, XdY$ and the gauge invariant interaction being 
given by some polynomial of the fields. 
(General arguments supporting that this is the only class 
of examples have been put forward in reference  \cite{Dayi:1993fk}.)
All of these theories are of {\em topological} nature.

   If some of the classical 
   equations of motion are algebraic (and linear 
   in the basic fields), as it is the case in examples 2, 3b, 3c, then they
   imply all other equations of motion by application of the covariant
   derivative. Moreover, elimination of fields by some algebraic equations
   of motion then reduces the first order actions
to second order actions analogous to TYM, i.e.  
 topological models of Witten-type.

\subsubsection{Other models and possible generalizations}

Some examples related to $2$-dimensional gravity
  have been studied in references \cite{Baulieu:1996ep, vilar}. 
  In this case, the components of the space-time metric 
  are viewed as gauge potentials associated to the invariance under
  general coordinate transformations.

  A further extension of the formalism is the following. 
  Consider the case of an abelian gauge group. 
The pairs of potentials $(\tilde X_p , \tilde Y_{d-p-1})$ 
 can be  generalized  to mixed dual pairs 
$(\tilde X_p , \tilde d \tilde Y_{d-p-2})$ and 
$(\tilde d \tilde X_p , \tilde Y_{d-p-2})$ 
each of which involves an abelian potential and an abelian 
field strength. Such
pairs appear in the transgression construction of $(d+1)$-dimensional
topological field theories  from  $d$-dimensional
topological models \cite{baura}. 

A different generalization of the algebraic formalism 
consists of introducing incomplete ladders and deformations of the 
operator $\tilde d \equiv d + s$ \cite{carvalho}. This approach allows to
discuss cohomological aspects of Yang-Mills-type theories
or supersymmetric extensions thereof \cite{carvalho, sasa}. 

Finally, we note that Yang-Mills theories can be formulated 
in terms of a first order action by 
deforming a BF model \cite{cotta}.
Thus, the algebraic formalism discussed here
should also be useful for describing 
these (non-topological) field theories.

\subsubsection{Master equation and gauge-fixing}

In this section, we summarize the general recipe for deriving
the vertex functional of the theory on a generic space-time 
manifold ${\cal M} _d$. 

The minimal actions $S_{\rm min}[\Phi^a ;\Phi^*_a]$ 
presented in section 5.1.3 
(which have been obtained 
 from a horizontality condition and cocycle condition) 
satisfy the BV master equation (\ref{master}).
For each of these models, 
 the gauge degrees of freedom have to be fixed by virtue of  
some gauge-fixing conditions ${\cal F}_\alpha$. The latter 
are implemented in the action by introducing a 
gauge-fermion $\Psi_{\rm gf}$ of ghost-number
$-1$  depending on antighost fields $\bar C^\alpha$: 
\bea
\Psi_{\rm gf} [\Phi^A] =  \intM{d} 
\tr \{ \bar C^\alpha {\cal F}_\alpha \} 
\, .
\label{gaugefermion}  
\eea 
The $s$-variation of 
$\bar C^\alpha$ yields the multiplier field 
$\Pi^\alpha$,
\be
s \bar C^\alpha=\Pi^\alpha 
\quad , \quad s\Pi^\alpha=0
\ee
and the corresponding antifields are assumed to 
transform ``the other way round":
\be
s \Pi^*_\alpha 
= (-1)^{ (d+1) (|\bar C ^\alpha | + 1) } \bar C^* _\alpha
\quad , \quad s \bar C^*_\alpha=0 
  \, .
\ee
These trivial (``contractible'') BRST-doublets, which do not contribute to
the physical content of the theory, are taken into account by adding 
a contribution 
\bea 
S_{\rm aux}[\bar C^*_\alpha,\Pi^\alpha]= - \intM{d} 
\tr \left\{\bar C^*_\alpha \Pi^\alpha\right\}
\label{actionaux} 
\eea
to the action $S_{\rm min}$. 
The resulting {\em non-minimal action} depends on the 
fields $(\Phi ^A ) = (\Phi^a, \bar C^{\alpha} , \Pi ^{\alpha})$ 
and the corresponding antifields
$(\Phi ^*_A ) = (\Phi ^* _a, \bar C ^* _{\alpha} , \Pi ^* _{\alpha})$: 
\bea 
S_{\rm nm}[\Phi^A,\Phi_A^*]
= 
S_{\rm min}[\Phi^a,\Phi^*_a]+S_{\rm aux}[\bar
C^*_\alpha,\Pi^\alpha]
\, . 
\label{actionnm}
\eea
It still solves the master equation. 

After elimination of the antifields according to prescription (\ref{exa}), 
one obtains the {\em vertex functional} 
\bea
        \Sigma[\Phi^A,\rho_A] = 
  \left.S_{\rm nm}\right|_{\Phi^*_A}
	=S_{\rm inv}+S_{\rm gf}+S_{\rm ext} +S_{\rm mod} 
	\, ,
\label{gamm}
\eea
where $S_{\rm gf} = s\Psi _{\rm gf}$ denotes 
the gauge-fixing part for the classical, gauge invariant action
$S_{\rm inv}$ 
and where $S_{\rm ext}$ represents the linear coupling of 
the external sources $\rho_A$ to the $s$-variations of the fields
$\Phi^A$.

\subsection{VSUSY}

In the following, we sketch the general procedure for obtaining 
the VSUSY-variations of all fields and antifields
on ${\cal M}_d = {\bf R}^d$. 

A $p$-form gauge potential $X_p$ generally admits 
a hierarchy of ghosts $X^1_{p-1}, X^2_{p-2},...$
and the gauge-fixing of the 
corresponding symmetries leads to analogous hierarchies of antighosts. 
All of these fields can be organized in a {\em BV-pyramid}
culminating in $X_p$, see Table \ref{pyramidX}.

\begin{table}[b]
\begin{center}
\begin{tabular}{ccccccccccccc} 
        &&&&& $X_p$& \\
        &&&&$\bar c_{p-1}^{\, -1}$&& $X_{p-1}^1$ & \\
        &&&$\bar c_{p-2}^{\, -2}$&&$\bar c_{p-2}$&&$X_{p-2}^2$  \\
        &&$\bar c_{p-3}^{\, -3}$&&$\bar c_{p-3}^{\, -1}$&&$\bar
	c_{p-3}^{\, 1} $&&$X_{p-3}^3$ \\  
	&$\bar c_{p-4}^{\, -4}$&&$\bar
	c_{p-4}^{\, -2}$&&$\bar c_{p-4}$&&$\bar c_{p-4}^{\, 2}
	$&&$X_{p-4}^4$
	\\
	\ldots &&&&&&&&&& \ldots
\end{tabular}
\captionit{$X_p$-pyramid}
\label{pyramidX} 
\end{center}
\end{table}

For a ladder $\tilde \Omega$, the VSUSY-variations are postulated to
be given by
\be
\label{gtl}
\del \tilde \Omega =\id  \tilde \Omega
\ ,  
\ee
i.e. VSUSY climbs the ladder from the highest ghost-number to the lowest
one. The {\em variations of the classical fields, the ghosts
and the associated antifields} follow directly from (\ref{gtl})
by choosing 
$ \tilde \Omega =  \tilde X_p, \,  \tilde Y_{d-p-1}$. 
As noted after eq.(\ref{dst}), the antifields transform in the other
direction than the fields do.

Next, we consider the {\em antighosts with negative ghost-number}, i.e.
those located on the left half  of the BV-pyramid, i.e. 
$\bar c_{p-1}^{\, -1}, \bar c_{p-2}^{\, -2}, \bar c_{p-3}^{\, -1},
\ldots$.
Their variations follow from the arguments preceding 
equations (\ref{vam}). Those of the {\em associated antifields}
are inferred from the general guideline that antifields 
transform in the other direction than fields do, i.e.
$\bar C^{\alpha} \stackrel{\del \,}{\to} \bar C^{\beta}$ implies
$\bar C^*_{\beta} \stackrel{\del \,}{\to} \bar C^*_{\alpha}$, 
see eq.(\ref{last}). 

All of the remaining {\em antighosts} have {\em positive} 
(more precisely non-negative)  {\em ghost-number}.
Those which have the same total degree can be gathered 
in ladders which correspond to the diagonals on the right half of the 
pyramid :
\bea 
\tilde {\bar c}_{p-2} \!\!\!&=&\!\!\! \bar c_{p-2}+\bar c_{p-3}^{\, 1} +\ldots
+\bar c^{\, p-2}
\label{lad}
\\
\tilde {\bar c}_{p-4} \!\!\!&=&\!\!\! \bar c_{p-4}+\bar c_{p-5}^{\, 1}
+\ldots +\bar c^{\, p-4}
\quad , \quad ...  
\nm 
    \eea
These generalized forms are incomplete since 
they only involve components with positive 
ghost-number\footnote{Yet, 
the BV-pyramid does not involve the complete ladder $\tilde X _p$
either, but only those components which have positive ghost-number.}.
Similarly, the {\em antifields associated to the antighosts} (\ref{lad}) 
can be collected into ladders containing only components 
with  negative ghost-number :
\be
(\tilde {\bar c}_{p-2})^* =
(\bar c^{\, p-2})^*+\ldots+(\bar c_{p-2})^* 
\quad, \quad ...
\ee
The transformation law (\ref{gtl}) is now postulated
for all of these 
ladders, i.e. for $\tilde \Omega = \tilde {\bar c}_{p-2n},
(\tilde {\bar c}_{p-2n})^*$ with $n=1,2,...$. 

The {\em $\del$-variations of the multipliers} 
$\Pi^\alpha = s \bar C^\alpha$ follow from the variations 
of the $\bar C^\alpha$ by requiring the VSUSY-algebra 
$[s, \del ] = \Li$ to be satisfied : 
\[
\del \Pi^\alpha = \del s \bar C^\alpha = 
\Li \bar C^\alpha - s (\del \bar C^\alpha ) 
\ . 
\]
The antifields $(\bar C^*_\alpha, \Pi^*_\alpha)$
associated to the BRST-doublets $(\bar C^\alpha,\Pi^\alpha)$ 
again transform ``the other way round".

\section{Conclusion}

As is well-known, the BV-formalism 
represents a systematic procedure 
for constructing an $s$-invariant action 
in the case of a gauge algebra
which is reducible and/or only valid on-shell. 
The $s$-operator of the BV-setting
 is nothing but the linearized Slavnov-Taylor operator.  
If the symmetry algebra is only valid on-shell, 
antifields appear in the $s$-variations 
and the solution $S_{\rm nm}$ 
of the Slavnov-Taylor identity
involves terms that are 
quadratic (or of higher oder) in the antifields. 

The algebraic framework for the BV-formalism 
on which we elaborated here, represents 
an elegant procedure for constructing solutions of the 
Slavnov-Taylor identity for {\em topological 
models of Schwarz-type} as defined in various dimensions. In 
particular, it allows to obtain quite straightforwardly 
the VSUSY-transformations which are most useful for 
dealing with the quantum version of these  theories. 

As emphasized in section 5, 
  {\em topological models of Witten-type} can also 
be introduced along these lines. However, their 
BRST- and VSUSY-algebras close off-shell in the standard BRST-approach
and therefore the introduction of antifields is not useful 
for their description 
\cite{Brandhuber:1994uf, Gieres:2000pv, Gieres:2000dr}.

Our discussion of VSUSY for topological models defined 
on flat space-time can be generalized to 
generic manifolds by incorporating VSUSY in the 
$s$-operation: this leads to an exact rather than 
a broken Ward identity and it proves to be useful 
for discussing the relationship between topological models 
and gravity \cite{cgp}.  

To conclude, we note that it would be interesting to gain a 
  deeper geometric understanding \cite{schwarz}
of the algebraic construction of topological models 
summarized in section 5. (Presumably the field theoretic formulation 
of references \cite{carval, sasa} provides the appropriate 
framework for this endeavor.) 
Such an insight should explain more fully  
why highly non-trivial solutions of the master equation 
can be obtained from a simple algebraic procedure.

\newpage 

\vskip 1.2truecm
 
{\bf \Large Acknowledgments}
 
\vspace{3mm}

%\nopagebreak 
 
F.G. wishes to thank M.~Schweda and 
all the members 
of the Institut f\"ur Theoretische Physik of the 
Technical University of Vienna
for the warm hospitality extended to him
during part of the present work. 
He also acknowledges enlightening discussions with 
O.~Piguet,  F.~Delduc and S.~Wolf.

\newpage

\providecommand{\href}[2]{#2}\begingroup\raggedright\endgroup

\end{document}